\documentclass{article}

\usepackage{natbib}
\usepackage[dvips]{graphicx}
\usepackage{url}
\usepackage{subcaption}

\begin{document}

\newcommand{\vect}[1]{\ensuremath{\vec #1}}
\newcommand{\italics}[1]{{\it #1}}

\title{PC Proxy: a Method for Dynamical Tracer Reconstruction}

\author{Peter Mills\\\textit{peteymills@hotmail.com}}

\maketitle

\pagestyle{myheadings}
\markright{P. Mills (2018)  \textit{Environ. Fluid Mech.} \textbf{18} (6): 1533-1558}

\section*{Abstract}

A detailed development of the principal component proxy method of dynamical
tracer reconstruction is presented, including error analysis.
The method works by correlating the largest principal components of a matrix
representation of the transport dynamics with a set of sparse measurements. 
The Lyapunov spectrum was measured and used to quantify the lifetime of each principal component. 
The method was tested on the 500 K isentropic surface with stratospheric ozone 
concentration measurements from the Polar Aerosol and Ozone Measurement (POAM) III satellite instrument
during October and November 1998 and compared with the older proxy tracer method
which works by correlating measurements with a single other tracer or proxy. 
Using a 60 day integration time and five (5) principal components, cross
validation of globally reconstructed ozone and comparison with ozone sondes returned root-mean-square errors of 0.16 ppmv and 0.36 ppmv, respectively.
This compares favourably with the classic proxy tracer method in which a passive
tracer equivalent latitude field was used for the proxy and 
which returned RMS errors of 0.22 ppmv and 0.59 ppmv for cross-validation
and sonde validation respectively.
The method seems especially effective for shorter lived tracers and was far
more accurate than the classic method at predicting ozone concentration in the
Southern hemisphere over the same time period.
It is also more effective when reconstruction is performed over the entire
Earth rather than a single hemisphere allowing for seamless reconstruction of
global fields.

\subsection*{Keywords}
\textbf{transport dynamics, interpolation methods, satellite remote sensing, tracer advection, assimilation models, numerical analysis, transport modelling; ozone}

\pagebreak

\tableofcontents

\section{Introduction}

Satellite remote sensing instruments have become invaluable for monitoring
trace atmospheric constituents such as ozone.
One area in which they frequently fall short is density of coverage.
Short of comprehensive prognostic assimilation models, there are methods of
interpolating sparse measurements that nonetheless take into account the
atmospheric dynamics.
One such method was first introduced in \citet{Butchart_Remsberg1986}
and used in \citet{Randall_etal2002} to derive Northern hemisphere maps
of ozone from Polar Ozone and Aerosol Measurement (POAM) III data.
\italics{Proxy tracer analysis} works by correlating
measurements with another atmospheric variable, the proxy, that 
is conserved by the flow.
The two most common proxies are potential vorticity \citep{Hoskins_etal1985} 
or a passive tracer simulated over a long time scale \citep{Allen_Nakamura2003}.
In either case, it's standard practice to transform the proxy to an
area-conserving \italics{Lagrangian coordinate} called equivalent latitude
before performing the reconstruction.
This method is mainly useful for long-lived tracers such as ozone.

Here we introduce a new method of dynamical tracer reconstruction
called \italics{principal component proxy analysis} 
that is more appropriate for shorter-lived tracers
because it works on shorter time scales.
The method is summarized as follows: 
the passive tracer dynamics are represented as a matrix
which linearly maps the initial tracer configuration
to the final tracer configuration.
We term this matrix the \italics{transport map}.
The transport map is decomposed using
principal component analysis (PCA)--hence the name of the method--and the largest principal components are then fitted to the measurements.
Because the transport map is integrated over only a short time scale and 
because it represents a large number of possible tracer configurations, with the most
likely singled out through the decomposition, 
PC proxy should be better able to 
compensate for non-advective changes in concentration or sources
and sinks.

\section{Theory}

\subsection{Tracer dynamics}

The advection-diffusion equation is given as follows:
\begin{equation}
\frac{\partial q(\vect x, ~ t)}{\partial t} = \left \lbrace -\vect v(\vect x, ~t) \cdot \nabla + \nabla \cdot D \nabla \right \rbrace q(\vect x, ~ t)
\label{advection_diffusion}
\end{equation}
where $q$ is the tracer concentration, $\vect x$ is spatial position, 
$t$ is time, $\vect v$ is the fluid velocity, and $D$ is the diffusivity tensor.
There is no mass-conservation term, $q \nabla \cdot \vect v$,
either because $q$ is a volume-mixing ratio (vmr) or 
the fluid is incompressible ($\nabla \cdot \vect v = 0$).
Sources are neglected since this would make the model nonlinear.
See below.

In an Eulerian tracer simulation, the approximate value of $q$
is known only at discrete locations so can be represented as a vector,
$\vect q=\lbrace q_i \rbrace$.
The linear operator 
contained in the braces in Equation (\ref{advection_diffusion}) 
is approximated as a matrix, $A$, which is multiplied with $\vect q$:
\begin{equation}
\frac{\mathrm d \vect q}{\mathrm d t} = A(t) \vect q
\label{linear_ODE}
\end{equation}

As an illustration, consider a second-order finite difference scheme in one dimension:
\begin{eqnarray}
\frac{\partial q_i}{\partial t} & = & \frac{v_i (q_{i-1} - q_{i+1})}{2 \Delta x} +
	\frac{d (q_{i-1} + q_{i+1} - 2 q_i)}{\Delta x^2} 
	\label{finite_difference_diffusion}
	\\\nonumber
& = & \left (\frac{v_i}{2 \Delta x} + \frac{d}{\Delta x^2} \right ) q_{i-1} -
	\frac{2 d}{\Delta x^2} q_i 
	+ \left (- \frac{v_i}{2 \Delta x} + \frac{d}{\Delta x^2} \right ) q_{i+1} 
\end{eqnarray}
where $d$ is a scalar diffusion coefficient, $\Delta x$ is the grid spacing
and $v_i$ is the wind speed at the $i$th grid point.
Expressed as elements of the matrix, $A$:
\begin{eqnarray}
a_{i,i-1} & = & \frac{v_i}{2 \Delta x} + \frac{d}{\Delta x^2} \\\nonumber
	a_{i,i} & = & -\frac{2 d}{\Delta x^2} \\\nonumber
a_{i,i+1} & = & - \frac{v_i}{2 \Delta x} + \frac{d}{\Delta x^2}
\end{eqnarray}
The actual transport model used in this study is described in Section 
\ref{ctraj}.

To produce a general solution to Equation (\ref{linear_ODE}), 
we first substitute a matrix, $R$, for $\vect q$:
\begin{equation}
	\frac{\mathrm d R(t_0, t)}{\mathrm d t} = A(t_0+t) R(t_0, t)
\end{equation}
We will call the matrix $R$ the \italics{discrete transport map} or simply
the \italics{transport map}.

Unlike in most analyses, there are two parameters for the time:
the integration start time, $t_0$, and the integration time, $t$.
In this way, $R$ may be decomposed in terms of itself:
\begin{equation}
	R(t_0,~t_n-t_0) = R(t_n, \, \Delta t_n) R(t_{n-1},\,\Delta t_{n-1}) R(t_{n-2},\,\Delta t_{n-2}) \, ...~~ 
	R(t_0,\,\Delta t_0)
\label{matrix_soln_decomposition}
\end{equation}
where,
\begin{equation}
t_n=t_0+\sum_{i=0}^{n} \Delta t_i
\end{equation}
It follows that $R(t, 0)=I$, for any $t$, where $I$ is the identity matrix.

Given $R$, 
we can calculate $\vect q$ given $\vect q$ at any other time:
\begin{equation}
	\vect q(t) = R(t_0, t-t_0) \vect q_0
	\label{tracer_integration}
\end{equation}
where $\vect q_0 = \vect q(t_0)$ is the initial tracer configuration.

To make this solution fully analytical, although not realizable in practice,
we solve $R$ for an infinitessimal integration time using linear algebra:
\begin{equation}
	\lim_{\Delta t \rightarrow 0} R(t, \Delta t) = \exp \left [ \Delta t A(t) \right ]
\end{equation}
where $\exp$ is the matrix generalization of the exponential function.

\subsection{Principal component proxy}

Suppose we decompose a given transport map using singular value decomposition:
\begin{equation}
	R(t_0, ~ t_n - t_0) = U S V^T
	\label{SVD}
\end{equation}
where 
$t_n - t_0$ is the \italics{integration time},
$S$ is a diagonal matrix of \italics{singular values},
and both $U$ and $V$ are orthogonal matrices:
\begin{equation}
	U^T U = V^T V = I~,
\end{equation}
$U$ is the matrix of \italics{left singular vectors} and 
$V$ is the matrix of \italics{right singular vectors} \citep{Press_etal1992}.
This method of matrix decomposition is also known as \italics{principal component}
analysis or PCA, hence the name of the interpolation technique.
Singular vectors are increasingly being used in meteorology both to quantify the
predictability of a forecast and to generate perturbations for ensemble
forecasts \citep{Tang_etal2006}.

A good way to think of it is as a set of orthogonal initial conditions--the right singular vectors, $\lbrace \vect v^{(i)} \rbrace$--which, when the tracer dynamics are applied, 
map onto a set of orthogonal final conditions--the left singular vectors, $\lbrace \vect u^{(i)} \rbrace$--that have grown by respective factors $\lbrace s_i \rbrace$:
\begin{equation}
	R \vect v^{(i)} = s_i \vect u^{(i)}
\end{equation}
The superscript denotes the column number: 
this notation will be used throughout.
Also, the term PC or principal component will be used as a synonym for left singular vector.

The singular values are all positive and by convention are arranged from largest to smallest:
\begin{equation}
	s_i \le s_{i-1}
\end{equation}
The matrix can normally be reconstructed to a high degree of accuracy
using only a few of the largest singular values and vectors.
This also makes the problem tractable since a typical size for
the transport map might be $[(360\cdot181) \times (360\cdot181)] = [65160\times65160]$.

To correlate a set of sparse measurements, $\lbrace m_i \rbrace$,
with the top $k$ principal components, we first find interpolates at each
measurement location in the left singular vectors and then find a set
of coefficients, $\vect c$, that minimizes the mean-square error of the
interpolates versus the measurements.
In any real problem, the measurements are unlikely to occur at the same time,
so rather than using the left singular vectors, we use $R$ to advance the
right singular vectors to the same time as the measurement.
The time period during which measurements are admitted is the \italics{measurement window}.
While the difference between the integration start time, $t_0$,
and the center of the measurement window is the \italics{lead time}.

We have the following minimization problem:
\begin{equation}
\min_{\vect c} \sum_i \left \lbrace \sum_{j=1}^k c_j \vect w_i R(t_0, ~ t^{(i)}-t_0) \vect v^{(j)} - m_i \right \rbrace^2
	\label{PC_proxy_def}
\end{equation}
where $t^{(i)}$ is the time stamp of the $i$th measurement, $m_i$, 
and $\vect w_i$ is a vector of interpolation coefficients.
In this work, bilinear interpolation is used to interpolate measurement
locations. 
Since $R$ is only known at discrete time values, 
$R(t_0, ~t^{(i)}-t_0) \vect v^{(j)}$ 
is also approximated through linear interpolation.

Once the coefficients have been fitted, the tracer is reconstructed as follows:
\begin{equation}
	\vect q(t_n) \approx \sum_{i=1}^k c_i \vect u^{(i)}
\end{equation}

\subsection{Classic proxy tracer}

\label{classic}

Contrast the above description of principal component proxy tracer
analysis with the original proxy tracer technique, hereafter
referred to as ``classic'' proxy tracer.
In the earlier method, the measurements are simply correlated with another
tracer (the proxy): either a passive tracer that has been advected
continuously with periodic re-normalization \citep{Allen_Nakamura2003} 
or some other quantity that is conserved by the flow 
such as potential vorticity \citep{Randall_etal2002,Hoskins_etal1985}.
The regression is typically done to second-order and the proxy variable
converted to an area-based ``Lagrangian'' coordinate called
equivalent latitude \citep{Butchart_Remsberg1986}.

If $\vec \Phi=\lbrace \Phi_i \rbrace$ is the proxy field, 
then we have the following minimization problem:
\begin{equation}
  \min_{\vect c^\prime} \sum_i \left \lbrace \sum_{j=0}^N c^\prime_j \left [\vect w_i \cdot \vect \Phi(t_i) \right ]^j - m_i \right \rbrace^2
\end{equation}
where $\vect c^\prime$ are the regression coefficients 
and $N$ is the order of the method. 
The tracer is reconstructed:
\begin{equation}
	q_i(t) \approx \sum_{j=0}^N c^\prime_j \Phi_i^j(t)
\end{equation}
where $i$ in this case runs from 1 to the number of grid points in the
proxy field.

\subsection{Lyapunov exponents}

Suppose we have a system of ordinary differential equations (ODEs):
\begin{equation}
	\frac{\mathrm d \vect r}{\mathrm d t} = \vect f(\vect r, ~ t)
	\label{ODE}
\end{equation}
where $\vect r$ is a vector of dependent variables.
Linearize this about $\vect r$ using the \italics{tangent vector},
$\nabla_{\vect r} f$:
\begin{eqnarray}
\frac{\mathrm d}{\mathrm d t} (\vect r + \delta \vect r) & \approx & \vect f + 
	\nabla_{\vect r} \vect f \cdot \delta \vect x \\
	\frac{\mathrm d}{\mathrm d t} \delta \vect r & \approx & \nabla_{\vect r} \vect f \cdot \vect r
\end{eqnarray}
where $\delta \vect r$ is a vector of \italics{infinitessimal error vectors}.
Now define the \italics{tangent model}, $H$, such that:
\begin{equation} 
	\frac{\mathrm d}{\mathrm d t} H = \nabla_{\vect r} \vect f H \\
\end{equation}

A passive Eulerian tracer simulation is linear:
taking Equation (\ref{linear_ODE}) as our system of ODEs in
(\ref{ODE}) then setting $\vect r=\vect q$, 
we have, $\vect f(\vect q, ~t)=A(t) \vect q$,
while the tangent vector is given as:
\begin{equation}
	\nabla_{\vect q} \vect f = A
\end{equation}
Hence the transport matrix, $R$, is equivalent to the tangent model, $H$.

The Lyapunov exponents are defined as the logarithms of the time averages
of the singular values in the limit as time goes to infinity:
\begin{equation}
\lambda_i = \lim_{t \rightarrow \infty} \frac{1}{t} \log s_i;
~~~~~~~\lambda_{i-1} \le \lambda_i
\end{equation}
where $s_i$ is the $i$th singular value \citep{Ott1993}.
For most systems:
\begin{equation}
|\delta \vect r| \approx |\delta \vect r(0) | \exp(\lambda_i t)
\label{lambda1}
\end{equation}
That is, as $H$ is integrated forward, the largest singular value and
the largest singular vector will increasingly begin to dominate
\citep{Ott1993}.
The Lyapunov exponents can help us gauge the significance of each
singular vector at a given lead time.

\subsection{Special properties}

An important property of flow tracers is that the amount of substance is 
conserved:
\begin{equation}
\sum_i q_i = const.
\label{mass_conservation}
\end{equation}
The equation is exact if the simulation uses an equal area grid
and the fluid is incompressible ($\nabla \cdot \vect v=0$).
From this it follows:
\begin{eqnarray}
\sum_i r_{ij} & = & 1 
\label{columns_sum_to_one}\\
\sum_i a_{ij} & = & 0
\label{columns_sum_to_zero}
\end{eqnarray}
See Appendix \ref{mass_conservation_derivation} for the derivation.

All gridded Eulerian tracer simulations are by necessity diffusive. 
Given in addition the constraint above in (\ref{mass_conservation}),
we can also show that all the singular values are less-than-or-equal-to one:
\begin{equation}
	0 \le s_i \le 1
	\label{sv_lt_one}
\end{equation}
thus the Lyapunov exponents in turn will all be negative or zero.
Section \ref{Lyapunov_exponents} provides a numerical demonstration while 
Appendix \ref{Lyapunov_exponents_less_than_zero} gives the derivation.

\subsection{Error analysis}

A detailed error analysis can help us both to better understand the technique
and to chose the best parameters for a given interpolation.
Real flow tracers will have sources and sinks, thus we introduce a source
term, $\sigma$, to Equations (\ref{matrix_soln_decomposition})
and (\ref{tracer_integration}):
\begin{eqnarray}
	  \vect q(t_n) 
  & \approx & \vect \sigma_{n} + \nonumber \\
  & & R(t_{n-1}, ~ \Delta t_{n-1}) [\vect \sigma_{n-1} + \nonumber \\
  & & R(t_{n-2}, ~ \Delta t_{n-2}) [\vect \sigma_{n-2} + \nonumber \\
  & & \vdots \nonumber \\
  & & + R(t_1, ~ \Delta t_1) [\vect \sigma_1 + \nonumber \\
  & & R(t_0, ~ \Delta t_0) \vect q_0 ]...]]
\end{eqnarray}
where $\vect \sigma_i$ is the integrated source term for the $i$th time step.
Expanding:
\begin{equation}
\vect q(t_n) 
   = R(t_0,~t_n-t_0) \vect q_0 + \sum_{i=1}^{n} R(t_i,~t_n-t_i) \vect \sigma_i
  \label{sources_sinks}
\end{equation}

To get a handle on the error, we consider first a fully passive tracer 
(no sources or sinks) started with initial conditions $\vect q_0$.
Expanding this in terms of the right singular vectors with a set
of coefficients, $\vect c_0$:
\begin{equation}
	\vect q_0 = V \vect c_0
	\label{c0}
\end{equation}
means that the final tracer takes the following form:
\begin{equation}
	\vect q(t_n) = U S \vect c_0
\end{equation}
however we are only calculating the top $k$ singular vectors, so the
interpolation looks like this:
\begin{equation}
	\vect q(t_n) \approx \sum_{i=0}^k c_{0i} s_i \vect u^{(i)}
\end{equation}

The most significant source of error are the terms left out of the equation.
The smaller the remaining singular values, $\lbrace s_i|i=[k+1..n_p]\rbrace$,
the smaller the error, where $n_p$ is the total number of grid points in the tracer.
This is why the Lyapunov exponents are useful:
they tell us how fast the singular values shrink.

A tracer interpolated with all the singular vectors will take the form:
\begin{equation}
	\vect q(t_n) = U \vect c
	\label{full_interpolation}
\end{equation}
Substitution of (\ref{c0}) into (\ref{sources_sinks}) and comparison with
(\ref{full_interpolation}) produces
the following:
\begin{equation}
	\vect c = S \vect c_0 + U^T \sum_{i=1}^n R(t_i, t_n-t_i) \vect \sigma_i
\end{equation}
In other words, we need to project the integrated source terms onto the
singular vectors.
As before, the error is given by terms not included in the analysis:
\begin{equation}
	\vect \epsilon = \sum_{i=k+1}^{n_p} \vect u^{(i)} \left [s_i c_{0i}
	+ \vect u^{(i)} \cdot \sum_{j=1}^n R(t_j, t_n-t_j) \vect \sigma_j \right ]
	\label{error1}
\end{equation}
Note that the first occurrence of the singular vector, $\vect u^{(i)}$, does not
cancel out the second occurrence because the factor in square brackets is a scalar, i.e., order of operations matters.

The first term in Equation (\ref{error1}) sums the components of the initial tracer configuration that project onto the smaller singular vectors not included in the analysis. 
This term will be small because of the shrinking of singular values over time.
The second term are the components of the source terms projected onto the same singular vectors.
The less the source terms line up with the largest singular vectors, the
larger this source of error.

Equation (\ref{error1}) suggests two approaches to reducing the error.
The first is to reduce the size of $R(t_j, t_n-t_j)$ so that the source terms
grow as little as possible.
This would suggest that the measurement window should be in the middle
of the integration, i.e. the lead time is half the integration time.
The second is to make this factor as close as possible to $R(t_n, t_n-t_0)$
so that projection onto the singular vectors leaves the term $SV$ and
the leftover smaller components are shrunk by the singular values.
This would suggest that the measurement window should be towards the end
of the integration, i.e. the lead time is the same as the integration time.

To understand this last point, rewrite Equation (\ref{error1}) as follows:
\begin{equation}
	\vect \epsilon = \sum_{i=k+1}^{n_p} s_i \vect u^{(i)} \left [c_{0i}
	+ \vect v^{(i)} \cdot \sum_{j=1}^n R^{-1}(t_0, t_j-t_0) \vect \sigma_j \right ]
\end{equation}
Note that because of diffusion, 
a backwards integration is not equivalent to the inverse of the
forwards integration, but only approximately so, that is,
$R(t+\Delta t, -\Delta t) \approx R^{-1}(t, \Delta t)$, with the approximation
becoming worse as $\Delta t$ becomes larger.

Measurement error and fitting discrepancies can be treated in the same way 
as any other linear least squares problem. 
It stands to reason that having fewer measurements will magnify both
measurement errors and discrepancies generated by sources and sinks.
More measurements will allow the use of more singular vectors,
reducing both the error terms in (\ref{error1}).
In this paper we will take an empirical approach to the error analysis by
validating reconstructed tracer fields against actual measurements whenever
possible.

\section{Models and data}

\subsection{Tracer simulation}

\label{ctraj}

To generate the transport maps, the \textit{ctraj} software package is used (http://ctraj.sf.net) \citep{Mills2004,Mills2009}.  
The software is written in C++ and contain programs
for gridded, two-dimensional, semi-Lagrangian tracer advection on an 
azimuthally-equidistant-projected coordinate system:
\begin{eqnarray}
  x & = & r \cos \theta \\\nonumber
  y & = & r \sin \theta \\\nonumber
	r & = & R_E (\pi/2 - h \phi)
\end{eqnarray}
where $\theta$ is longitude, $\phi$ is latitude
and $h$ is the hemisphere in which the projection is
defined:
\begin{equation}
h = \left \lbrace \begin{array}{rl} 1; & \mathrm{North} \\ -1; & \mathrm{South} \end{array} \right .
\end{equation}
The resulting space has the following metric coefficients:
\begin{eqnarray}
\left (\frac{\mathrm ds}{\mathrm d x} \right )^2 & = & \frac{1}{r^2} \left [
	\frac{R_E^2}{r^2} \sin^2 \left (\frac{r}{R_E} \right ) 
	y^2 + x^2 \right ] \\ \nonumber
\left (\frac{\mathrm ds}{\mathrm d y} \right )^2 & = & \frac{1}{r^2} \left [
	\frac{R_E^2}{r^2} \sin^2 \left (\frac{r}{R_E} \right )
	x^2 + y^2 \right ]
	\label{metric_coef}
\end{eqnarray}
where $R_E$ is the radius of the Earth.
Two fields are advected simultaneously: one for the Northern hemisphere ($h=1$)
and one for the Southern hemisphere ($h=-1$), with equatorial crossings accounted for.

Because it is a semi-Lagrangian simulation, the factors, $\lbrace R(t_i,\Delta t) \rbrace$,
are output directly as sparse matrices by calculating the interpolation coefficients.
By storing the output as sparse matrices and not multiplying them through until
needed, it becomes possible to calculate the singular values and vectors through iterative
methods such as the Lanczoz method \citep{Golub_Van_Loan1996}.
The Arnoldi package (ARPACK) \citep{Lehoucq_Scott1996} is used to compute the
eigenvalues and eigenvectors needed for the SVD.

To calculate back-trajectories, a fourth-order Runge-Kutta integration scheme
was used with a 1.2 hour time-step.
Back trajectories were linearly interpolated after each 1-day, Eulerian time step.
Gridding in both hemispheres is 50 by 50, or 400km-,
3.6-degree-latitude-separation at the pole.

\subsection{Polar Ozone and Aerosol Measurement III instrument}

\begin{figure}
    \includegraphics[angle=-90,width=0.9\linewidth]{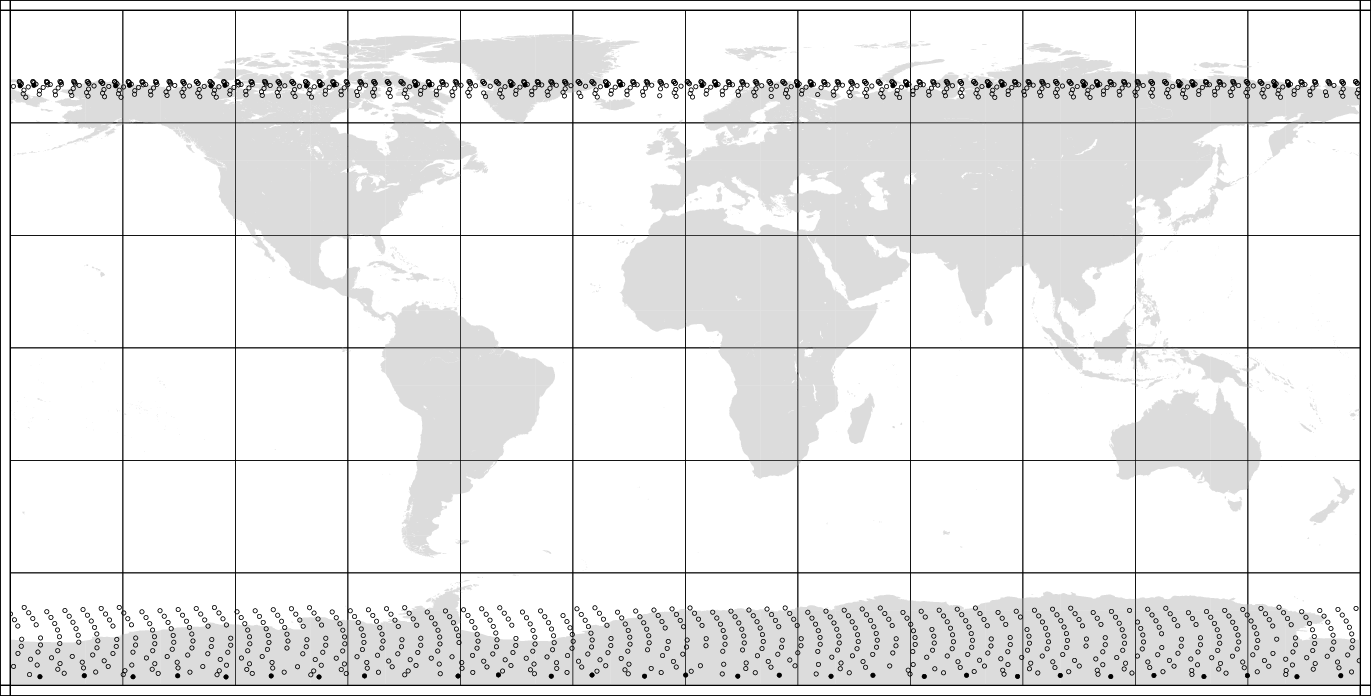}
    \caption{Locations of all POAM III measurements used in this study 
    between Sept. 25, 1998 and November 18, 1998. 
    Highlighted points are for the first time grid (two days worth of data) between Sept. 25, 1998 0:00 and Sept 27, 1998 0:00 UTC.}
    \label{POAM_loc}
\end{figure}

The Polar Ozone and Aerosol Measurement (POAM) III instrument was a solar-
occultation instrument mounted on the SPOT-4 sun-synchronous, low-earth-orbit
satellite \citep{Lucke_etal1999}.
Operating between March 1998 and December 2005,
it had nine channels in the visual and near infrared range.
Using optimal estimation \citep{Rodgers2000}, ozone profiles have been retrieved 
within a pair of narrow latitude bands in both the Arctic and Antarctic \citep{Lumpe_etal2002}.  
The instrument is capable of returning 28 or 29 measurements per day,
alternating between Northern and Southern hemisphere, however because of
a malfunction
in the instrument, it normally operates in only one or the other hemisphere for longer periods.  Therefore, we confine ourselves to earlier data,
October and November 1998, when more frequent and diverse measurements
are available.
The locations of all POAM measurements used in this study are plotted in
Figure \ref{POAM_loc}.

\subsection{National Center for Environmental Prediction reanalysis data}

The National Center for Environmental Prediction (NCEP) supplies, 
free-of-charge,
gridded (2.5 by 2.5 degrees longitude/latitude, 4 time daily), reanalyzed 
climate data starting in 1948 \citep{Kalnay_etal1996}.
Daily averaged wind and temperature data was used to drive the advection model.

\begin{figure}
  \begin{flushright}
    \includegraphics[angle=90,width=0.9\textwidth]{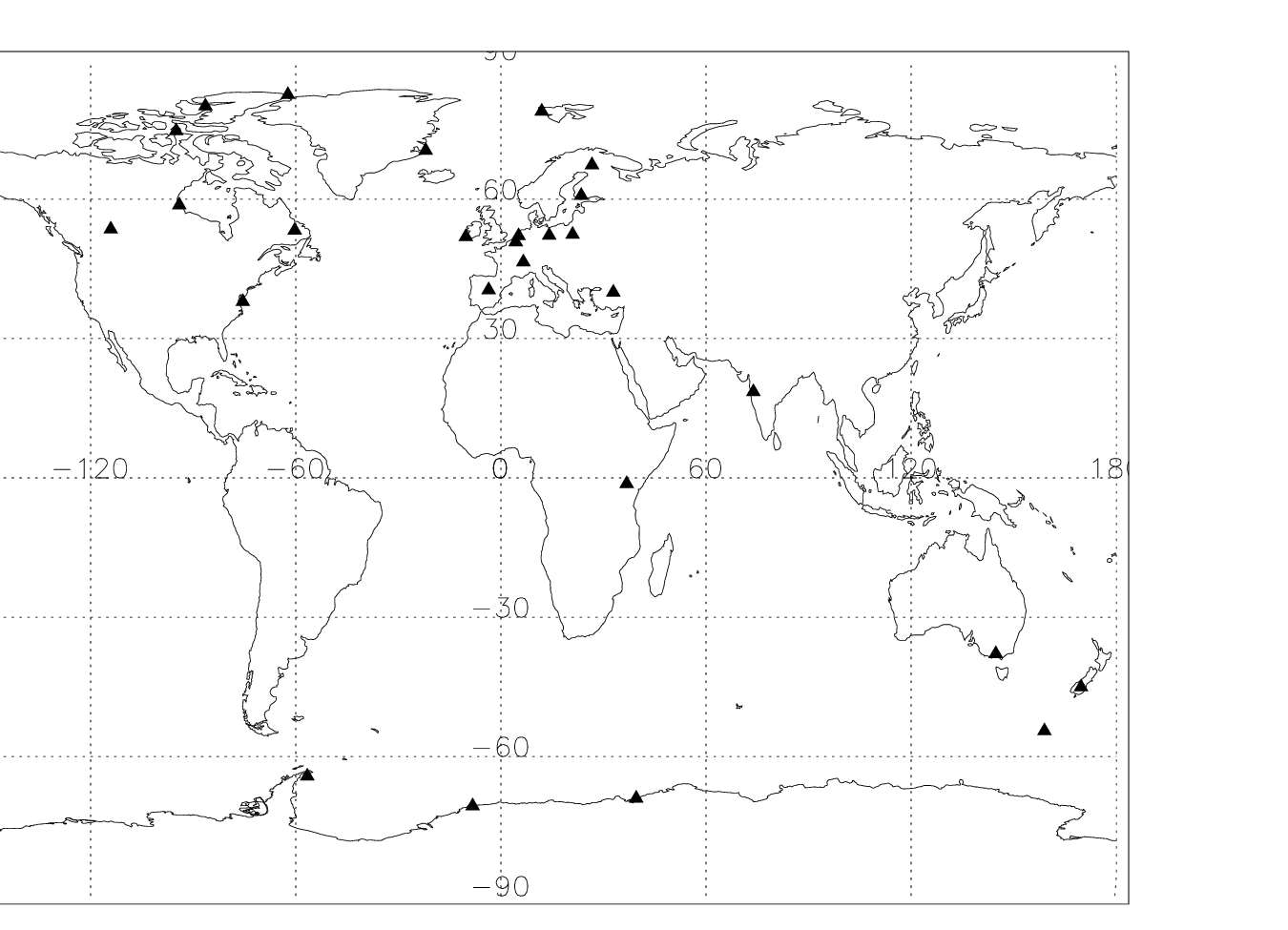}
    \caption{Locations of ozone sonde launch stations}
    \label{ozone_sonde_stations}
  \end{flushright}
\end{figure}

\subsection{Ozone sonde data}

The World Ozone and Ultraviolet Data Centre (WOUDC) collects ozone sonde data
from around the world \citep{Hare_etal2000}. A list of all contributors is available on the website:
\url{http://woudc.org}.
The data is archived by Environment Canada.
The location of all the launch stations used in the validation exercises is
shown in Figure \ref{ozone_sonde_stations}.
Sonde locations are a good match for the POAM III measurement locations
since they are primarily at high latitudes with few stations
towards the equator.

\section{Numerical preliminaries}

\begin{figure}
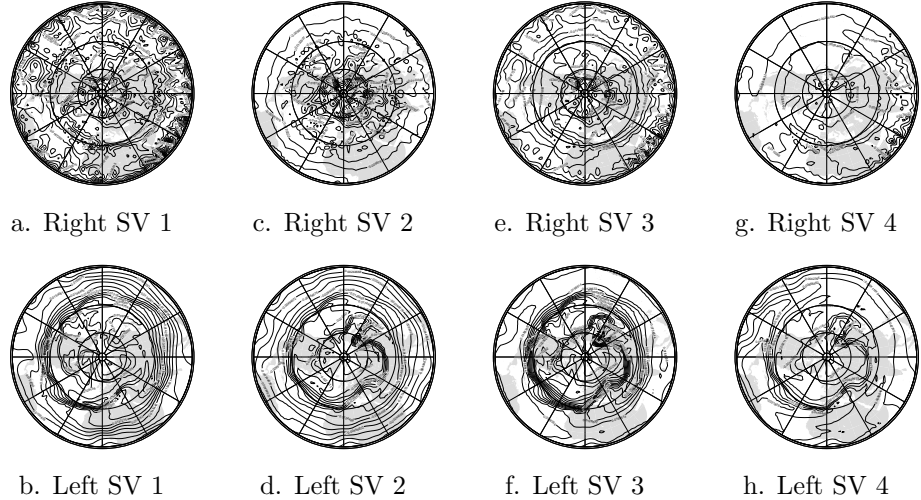

  \begin{tabular}{cccc}
    \includegraphics[angle=-90,width=0.23\textwidth]{svr4} &
    \includegraphics[angle=-90,width=0.23\textwidth]{svr3} &
    \includegraphics[angle=-90,width=0.23\textwidth]{svr2} &
    \includegraphics[angle=-90,width=0.23\textwidth]{svr1} \\
    a. Right SV 1 & c. Right SV 2 & e. Right SV 3 & g. Right SV 4 \\
    \includegraphics[angle=-90,width=0.23\textwidth]{svl4} &
    \includegraphics[angle=-90,width=0.23\textwidth]{svl3} &
    \includegraphics[angle=-90,width=0.23\textwidth]{svl2} &
    \includegraphics[angle=-90,width=0.23\textwidth]{svl1} \\
    b. Left SV 1 & d. Left SV 2 & f. Left SV 3 & h. Left SV 4
  \end{tabular}
  \caption{An example of the singular vectors derived from a discrete transport map.
  Sixty day integration started on August 1 1998 on the 500 K isentrop driven by NCEP wind fields.}
  \label{sample_SV}
\end{figure}

For reference, examples of singular vectors derived from the transport map,
$R(t, \Delta t)$, where $t$ is August 1, 1998 and $\Delta t$ is sixty days,
are shown in Figure \ref{sample_SV}.
If $R(t, \Delta t)=U S V^T$,
Figure \ref{sample_SV}a. shows the first right singular vector, $\vect v^{(1)}$,
while Figure \ref{sample_SV}b. shows the corresponding left singular vector,
$\vect u^{(1)}$;
Figure \ref{sample_SV}c. shows the second right singular vector, $\vect v^{(2)}$,
while Figure \ref{sample_SV}b. shows the corresponding left singular vector,
$\vect u^{(2)}$
and so on.
While the globally projected fields representing the right singular vectors 
appear to be quite similar and are indeed
quite strongly correlated, the abstract vectors from which they are derived
have negligible dot products.

\begin{figure}
\begin{center}
\includegraphics[width=0.9\textwidth]{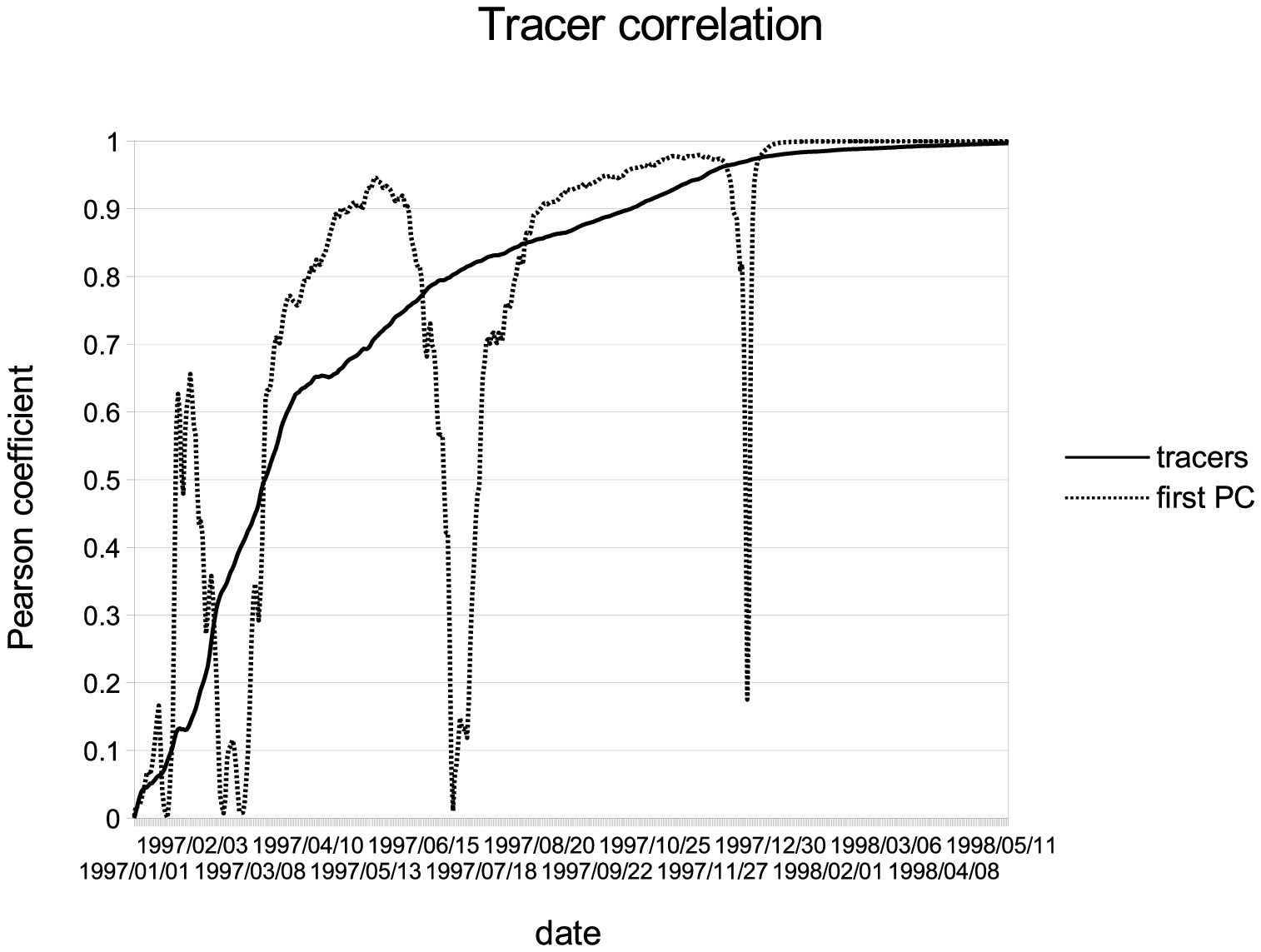}
\caption{The Pearson correlation coefficient over time of two differently-initialized,
	two-dimensional tracers
(broken line)--zonally symmetric and meridionally symmetric--and of
the zonally-symmetric-initialized tracer with the first principal component.
The simulation was driven with NCEP reanalysis 1 data on the 500 K isentropic
level with an Eulerian time-step of 20 hours and a Lagrangian time-step
of one 1.2 hours.}\label{tcorr}
\end{center}
\end{figure}

\begin{figure}
\begin{tabular}{ll}
a. Tracer & b. First PC\\
\includegraphics[angle=-90,width=0.45\textwidth]{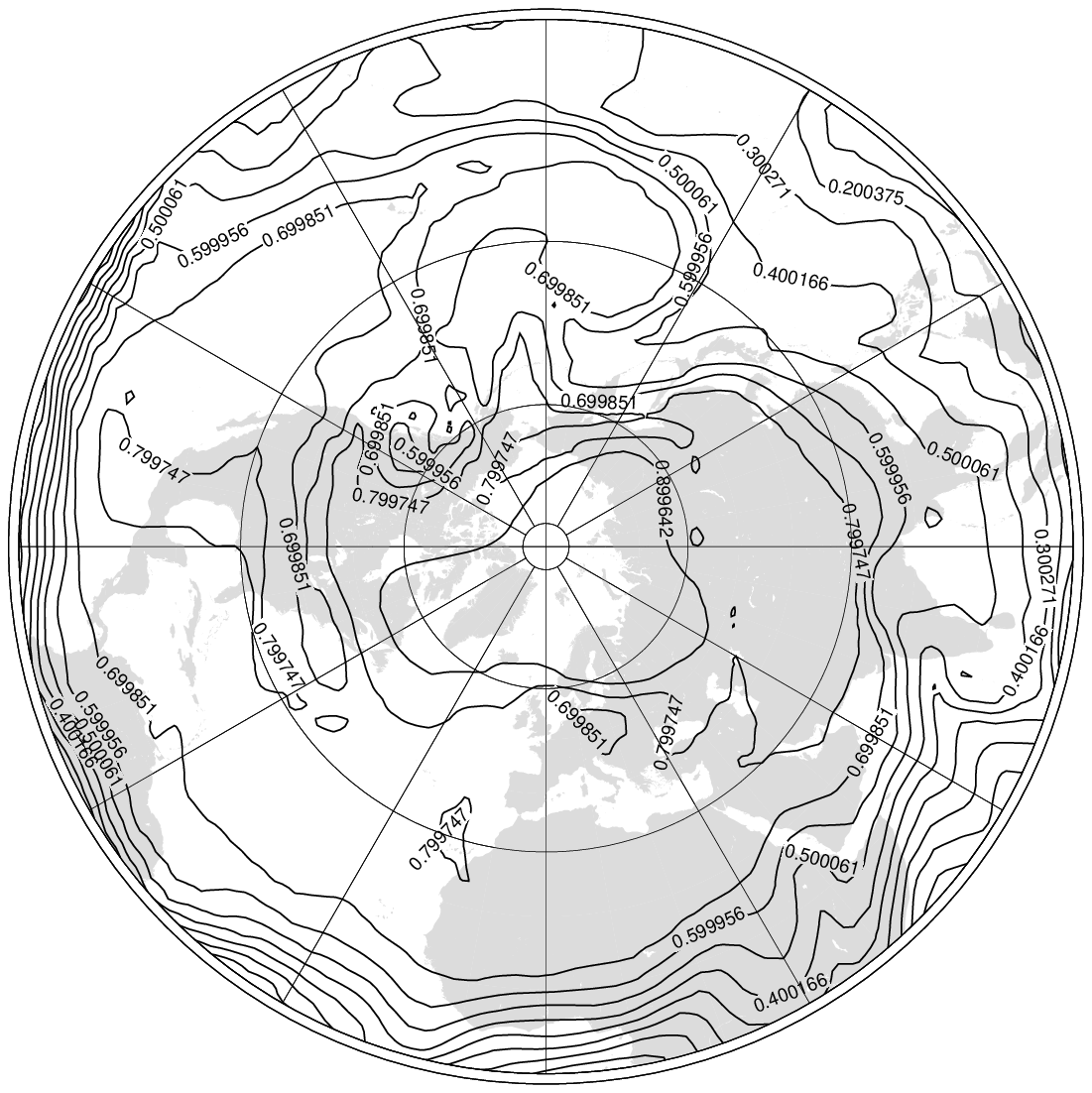} &
\includegraphics[angle=-90,width=0.45\textwidth]{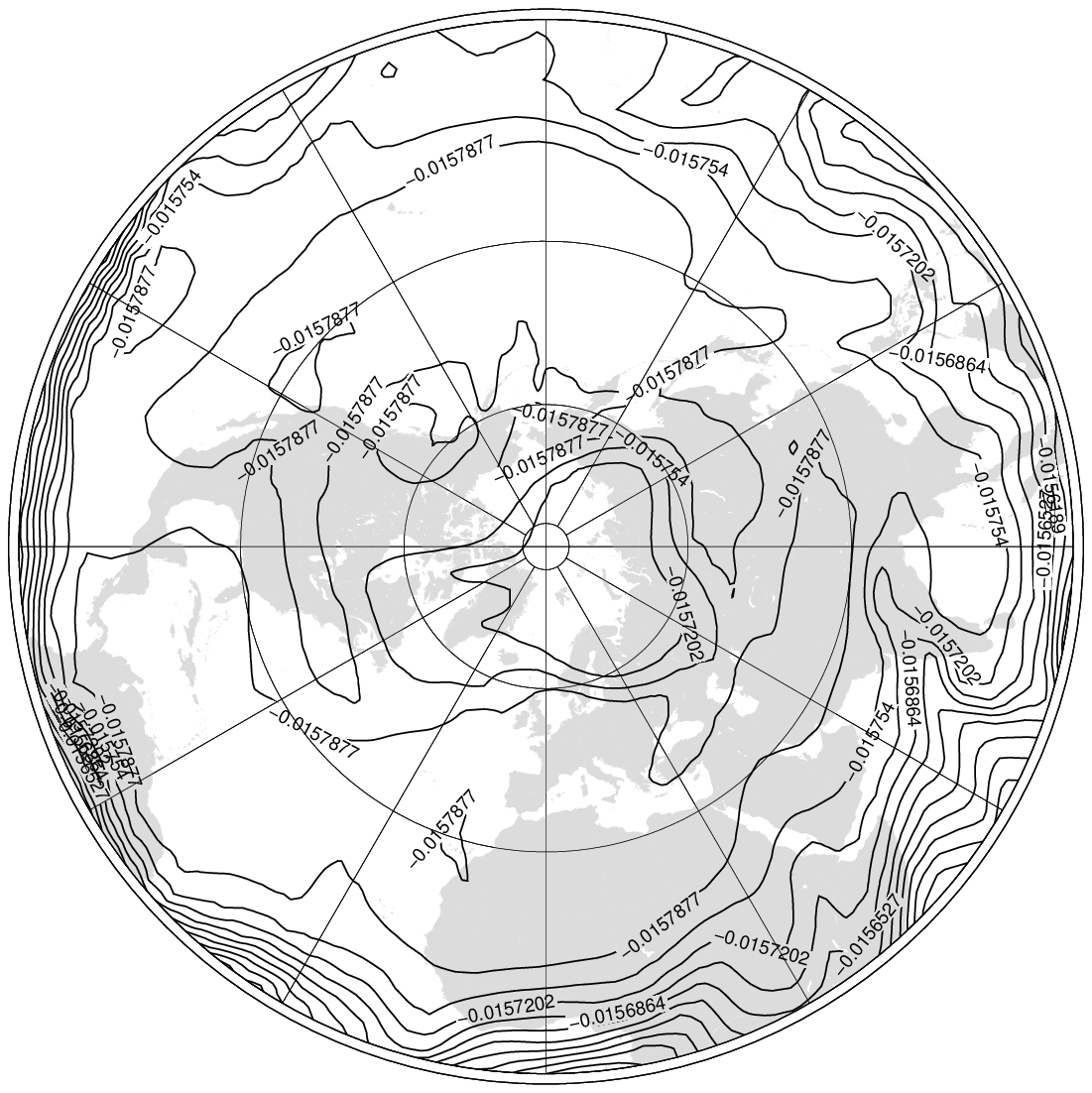}
\end{tabular}
\caption{Comparison of a simulated tracer (a.) and the first principal
component (b.) for the same time integration.
Integration started on January 1, 1997 and continued until December 1, 1997, a period of 334 days and was performed on the 500 K isentrop with NCEP wind fields.}\label{pc1}
\end{figure}

\subsection{Tracer correlation}

Two differently-initialized tracers, when integrated with the same
wind fields over a long time period, become correlated.
As discussed in Section \ref{classic},
this can be used to infer global fields of a long-lived tracer such as
ozone based on only a few sparse measurements 
\citep{Allen_Nakamura2003,Randall_etal2002}.
Figure \ref{tcorr} demonstrates this with the extreme example of an initially
zonally-symmetric tracer and an initially meridionally-symmetric,
two-dimensional tracer.
Tracers are passively advected with National Center for Environmental Prediction
(NCEP) reanalysis 1 data at the 500 K isentrop \citep{Kalnay_etal1996}.
The Pearson coefficient, weighted by grid size, is applied over the whole field at a single time-step.

We also plot the Pearson correlation coefficient of the first tracer with the largest singular
vector.  
We see that, because of Equation (\ref{lambda1}), they too become
correlated over time.
This at least partially explains the efficacy of the proxy tracer method.
The periodic dips in correlation are not yet understood.
A sample PC as compared with the tracer is shown in Figure \ref{pc1}.  

\subsection{Calculating Lyapunov exponents}

\label{Lyapunov_exponents}

\begin{figure}
\begin{center}
\includegraphics[width=0.9\textwidth]{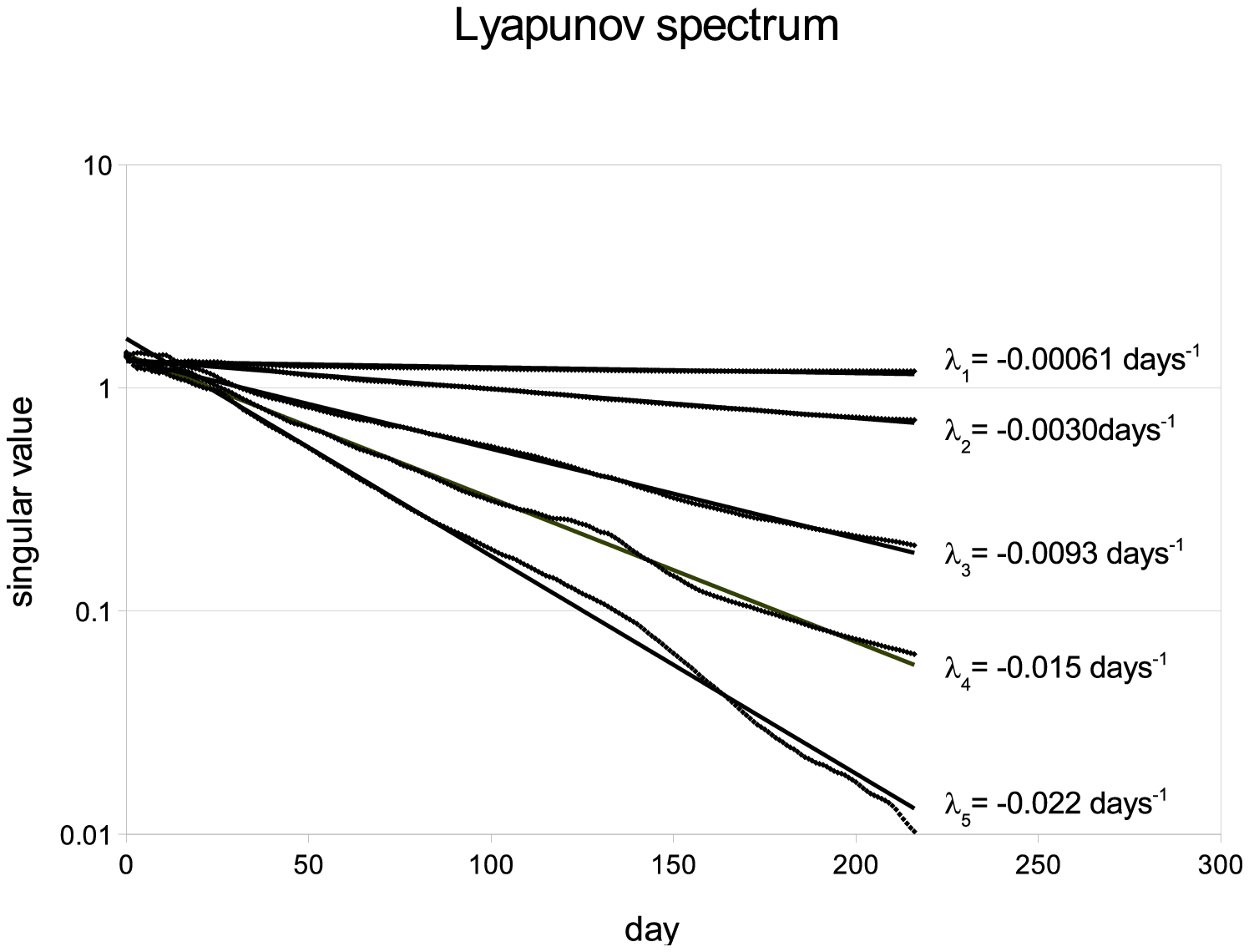}
\caption{Plot of the top five (5) singular-values of a semi-Lagrangian
tracer simulation over time.  Straight-line fits return the Lyapunov-spectrum.
The simulation was done on the 500 K isentropic level.}
\label{lyap_spec}
\end{center}
\end{figure}

Figure \ref{lyap_spec} plots the time evolution of the singular values.
From this we can calculate the Lyapunov spectrum by making straight line
fits of their logarithms.
While the resulting fields may develop into complex fractals \citep{Mills2009}
the Lyapunov spectrum shows that the tracer dynamics themselves 
are not truly chaotic (in fact cannot be), but are
only on the cusp: the largest singular value remains approximately constant.
It also shows how quickly the other singular vectors decay,
so that the largest will eventually dominate in accordance with
Equation (\ref{lambda1}).

All the Lyapunov exponents are less than zero, with the largest roughly zero, 
in agreement with the inequality in (\ref{sv_lt_one}),
even though Equation (\ref{mass_conservation}) is true only approximately.
In this case it is close enough: the smallest boxes on the
azimuthal equidistant grid will deviate from the largest by
a factor of only $2/\pi^2+1/2\approx0.7$ (see derivation in
Appendix \ref{equal_area})
while most real fluids are approximately non-divergent, especially
when considered over long time scales.

\section{Ozone reconstruction}

In this section we use PC proxy to reconstruct two-dimensional ozone fields 
from POAM III satellite data.
To perform the analysis, we need to choose an integration time, 
$t_n-t_0$, in Equation (\ref{SVD}), as well as a measurement window.
The integration time determines how long the tracer is advected 
before performing the SVD.
Measurements are selected from within the measurement window.
We also need to choose a lead time which determines how far from the
beginning of the integration, $t_0$, the measurement window is centered.
For this experiment, it was centered at the end of the integration period.
Placing the measurement window in the middle of the integration period
was found to work poorly.
The Lyapunov spectrum can help us select the number of singular vectors
as it shows how many remain significant at
a given lead time--see Figure \ref{lyap_spec}.

\begin{figure}
  \begin{tabular}{l}
    \includegraphics[angle=-90,width=0.9\linewidth]{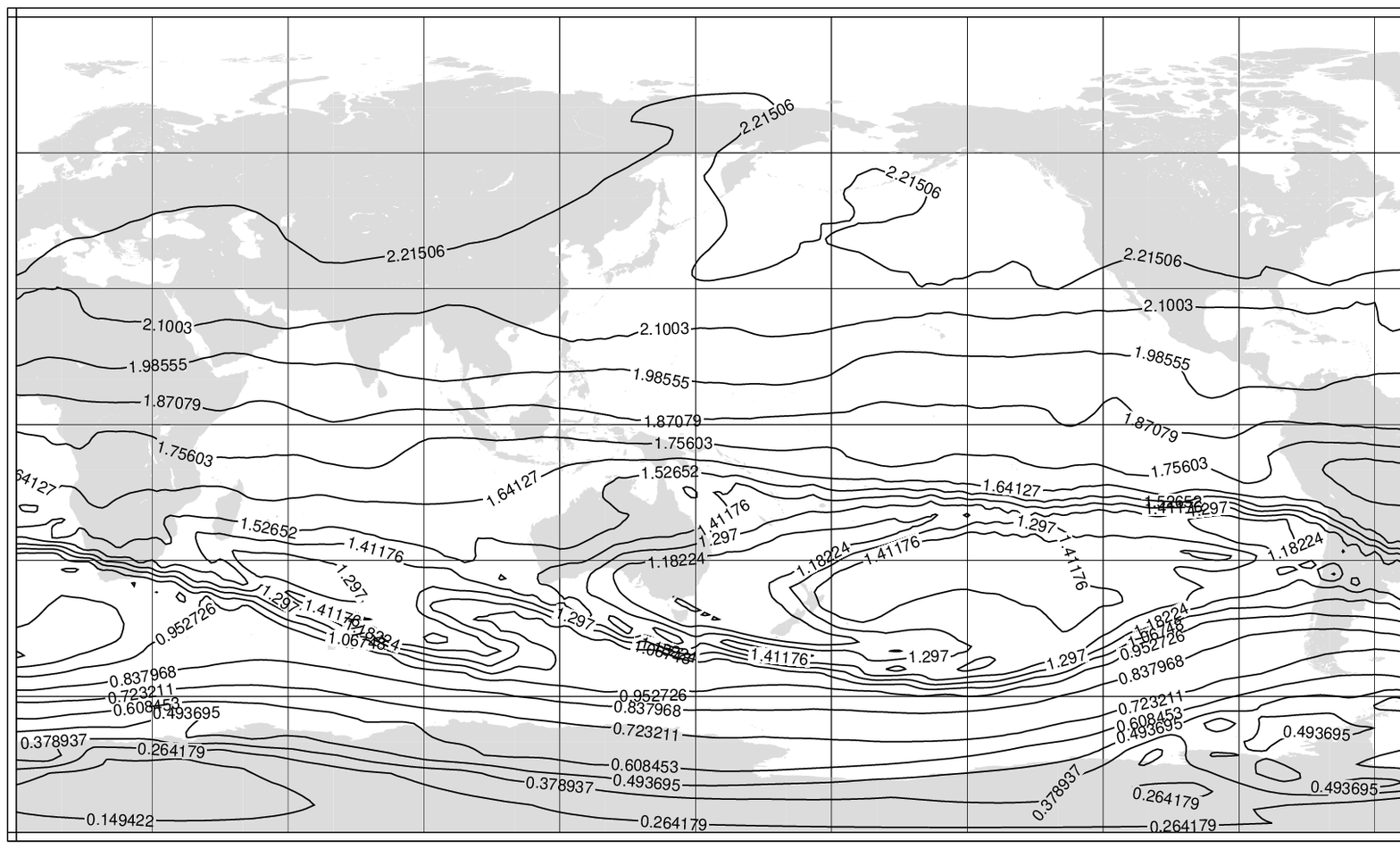} \\
    a. Classic proxy tracer \\
    \includegraphics[angle=-90,width=0.9\linewidth]{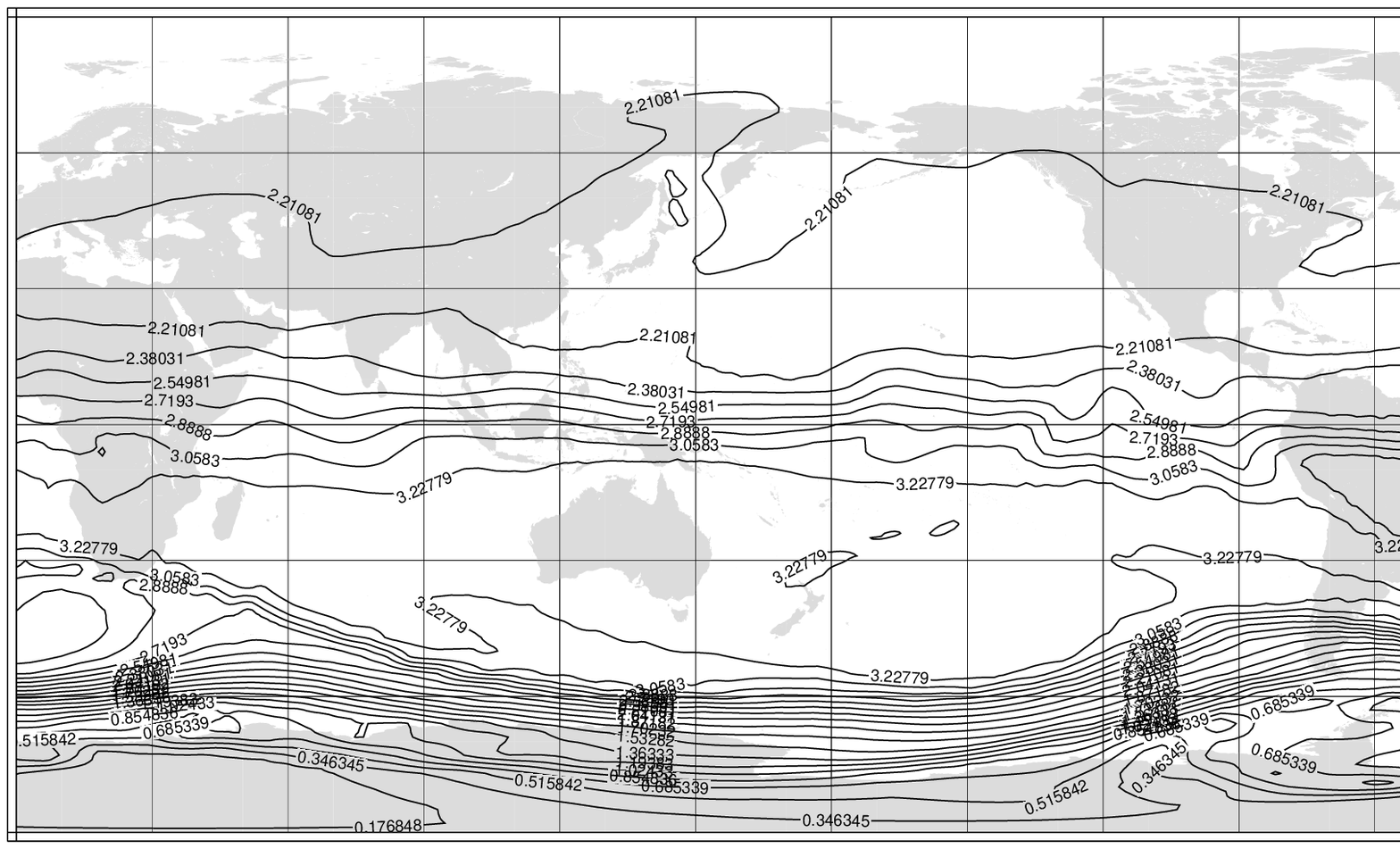} \\
    b. PC proxy
  \end{tabular}
  \caption{Sample globally reconstructed ozone fields on the 500 K isentrop for October 1, 1998. Units are parts-per-million by volume (ppmv).}
  \label{sample_O3_field}
\end{figure}

Ozone fields were reconstructed daily on the 500 K isentrop
between September 26 and November 17, 1998,
which is one of only a few periods in which POAM III was operating in both 
hemispheres simultaneously.
The tracer simulation was run at a 50 by 50 resolution or 400 by 400 km at
the poles with a 1.2 hour Runge-Kutta time step for the back-trajectories
and a 1 day Eulerian time step.
The integration time was 60 days with the same lead time
and five principal components were used unless otherwise noted.
The measurement window was two days.

An example of a reconstructed ozone field is shown in Figure \ref{sample_O3_field}
with Figure \ref{sample_O3_field}a. using the classic proxy tracer method
while Figure \ref{sample_O3_field}b. uses the PC proxy method.
Because POAM III data is confined to two rather narrow latitude bands near
the poles, values near the equator may not be that accurate.
Nonetheless, the author thought it important to test the method with global
reconstructions to see how well the PC proxy method can extrapolate
to areas where measurements are sparse or non-existant.
As we will see, PC proxy is not only more accurate than the classic method,
it can also be better at taking into account non-local information
when it does not become unstable due to over-fitting.

\subsection{Cross-validation}

\label{cross_validation}

\begin{figure}
  \centering
  \includegraphics[width=0.9\textwidth]{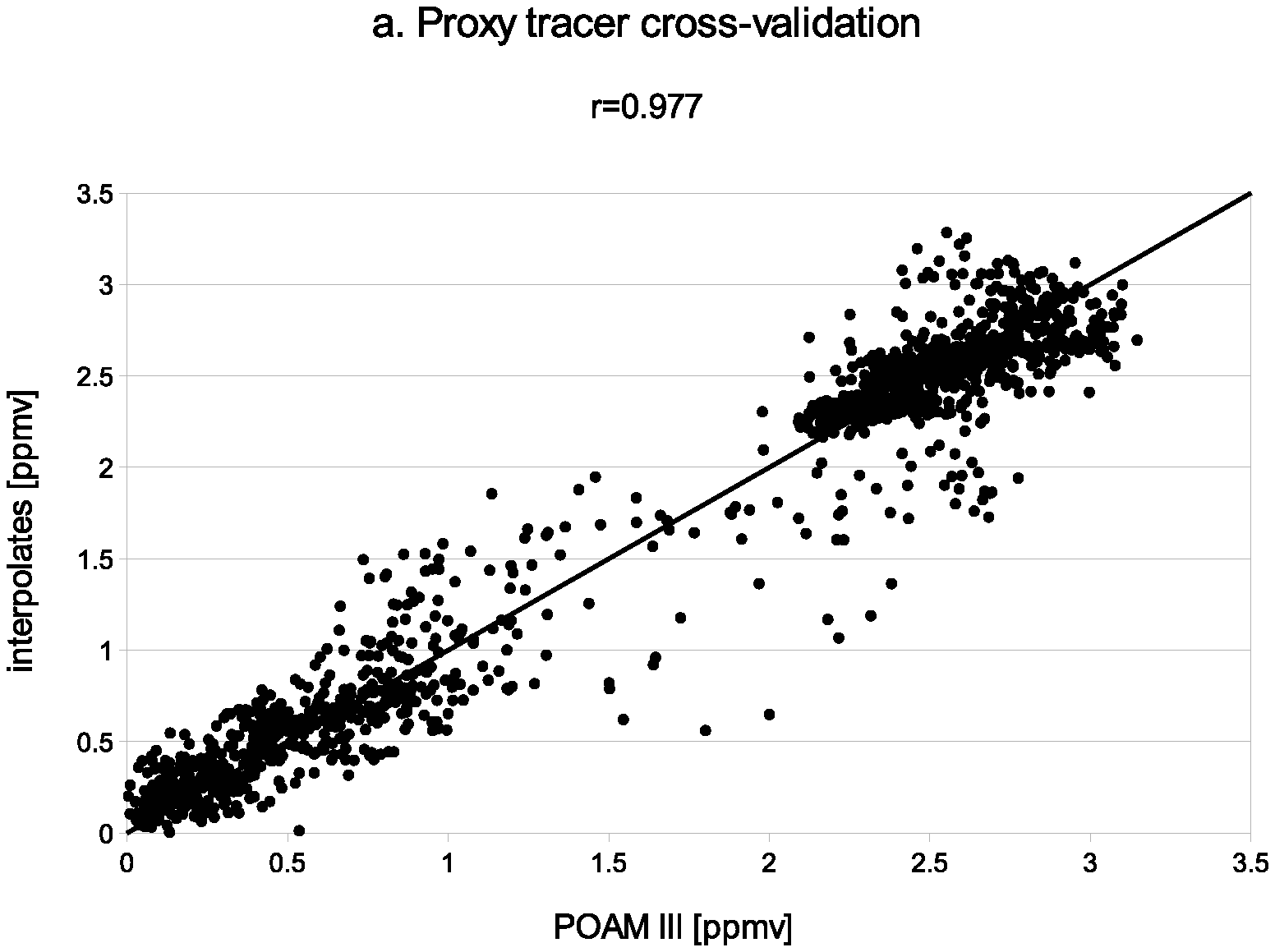}
  \includegraphics[width=0.9\textwidth]{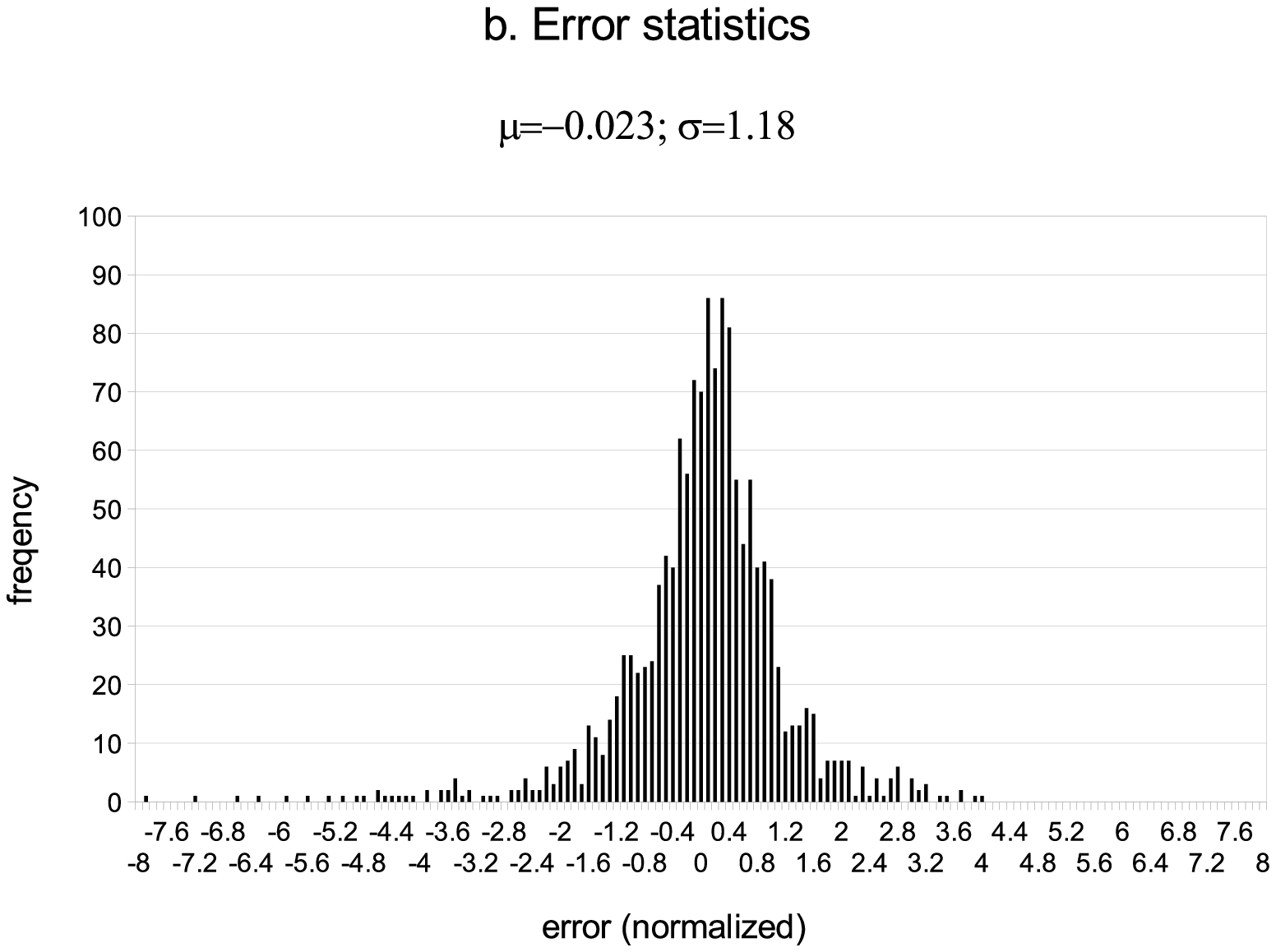}
  \caption{Cross validation of classic second-order proxy tracer ozone reconstruction. The reconstruction was done globally on the 500 K isentrop from POAM III data between September 26, 1998 and November 17, 1998 using a measurement window of 2 days.}
  \label{classic_cross_validation}
\end{figure}

\begin{figure}
  \centering
  \includegraphics[width=0.9\textwidth]{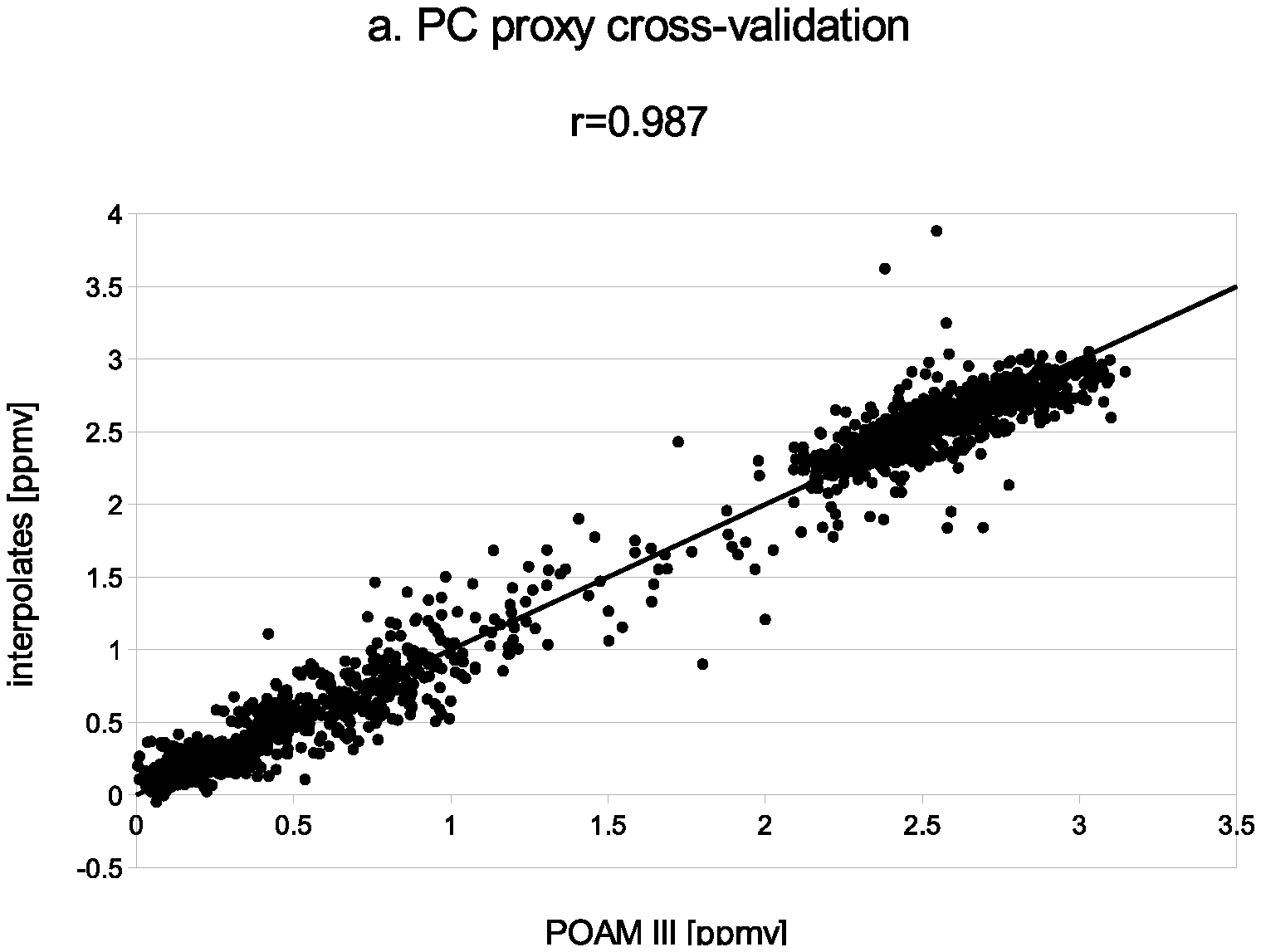}
  \includegraphics[width=0.9\textwidth]{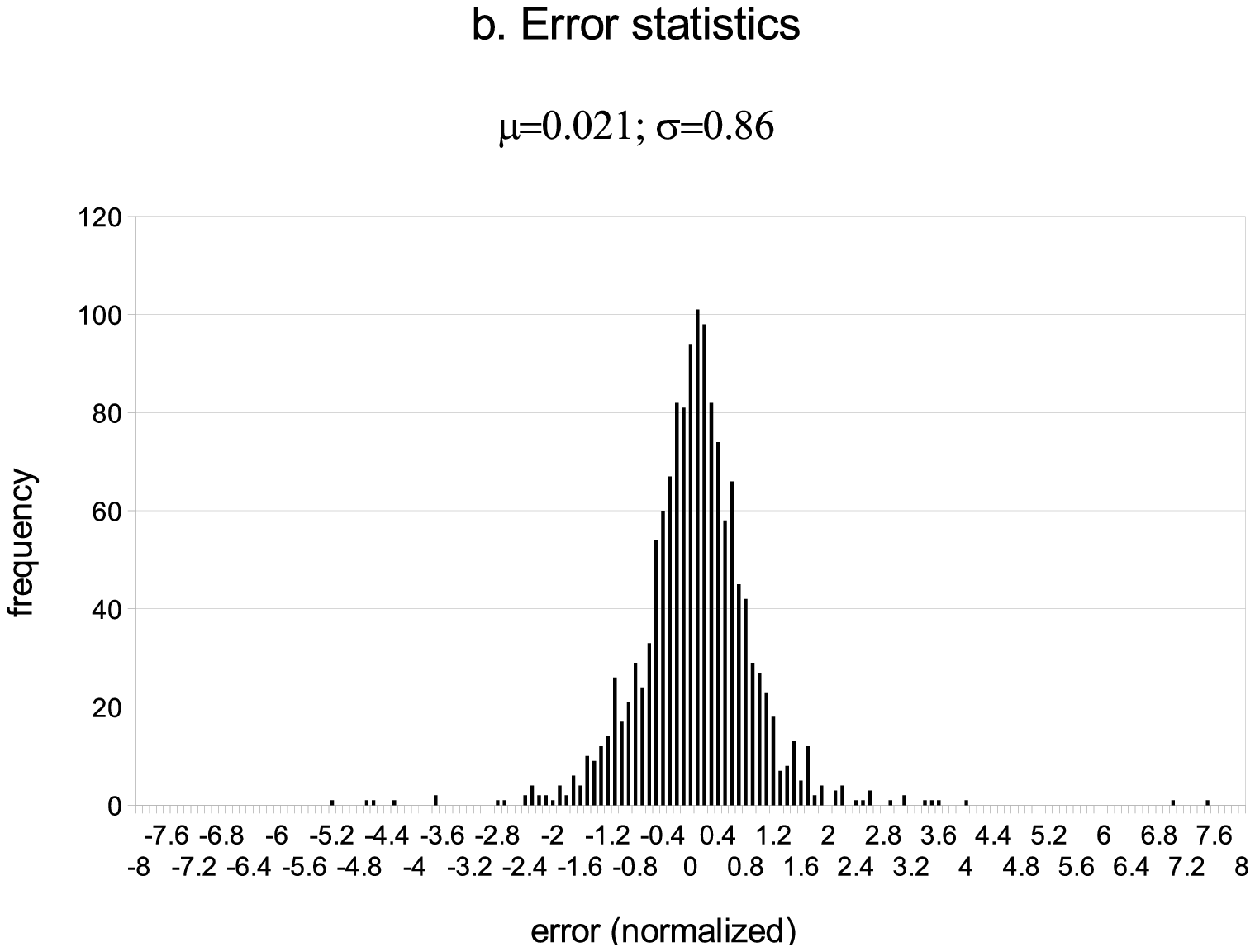}
  \caption{Cross validation of principal component proxy ozone reconstruction. The reconstruction was done globally on the 500 K isentrop from POAM III data between September 26, 1998 and November 17, 1998 using 5 principal components,  an integration time of 60 days, and a measurement window of 2 days.}
  \label{PC_cross_validation}
\end{figure}

\begin{figure}
  \centering
  \includegraphics[width=0.9\textwidth]{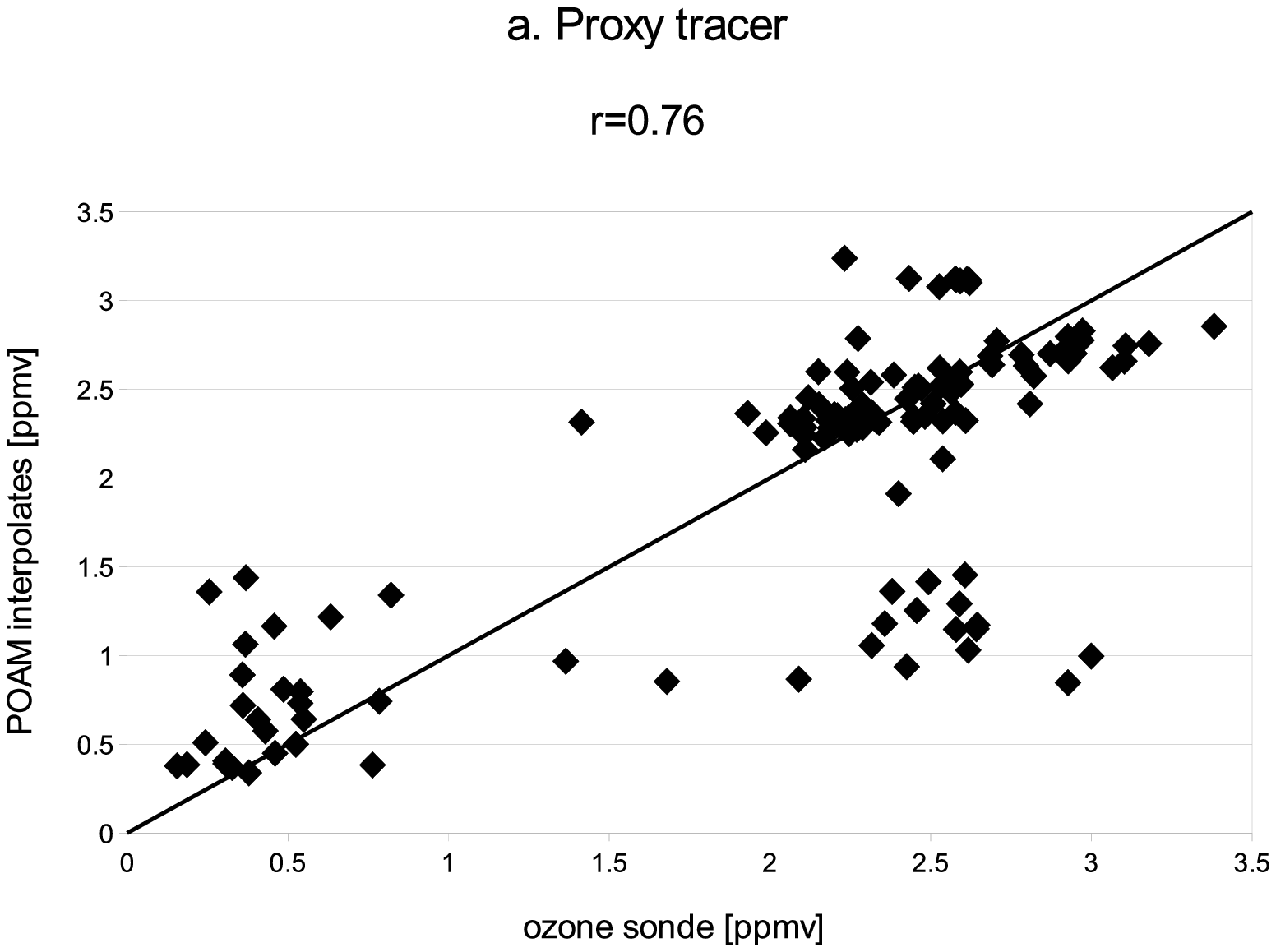}
  \includegraphics[width=0.9\textwidth]{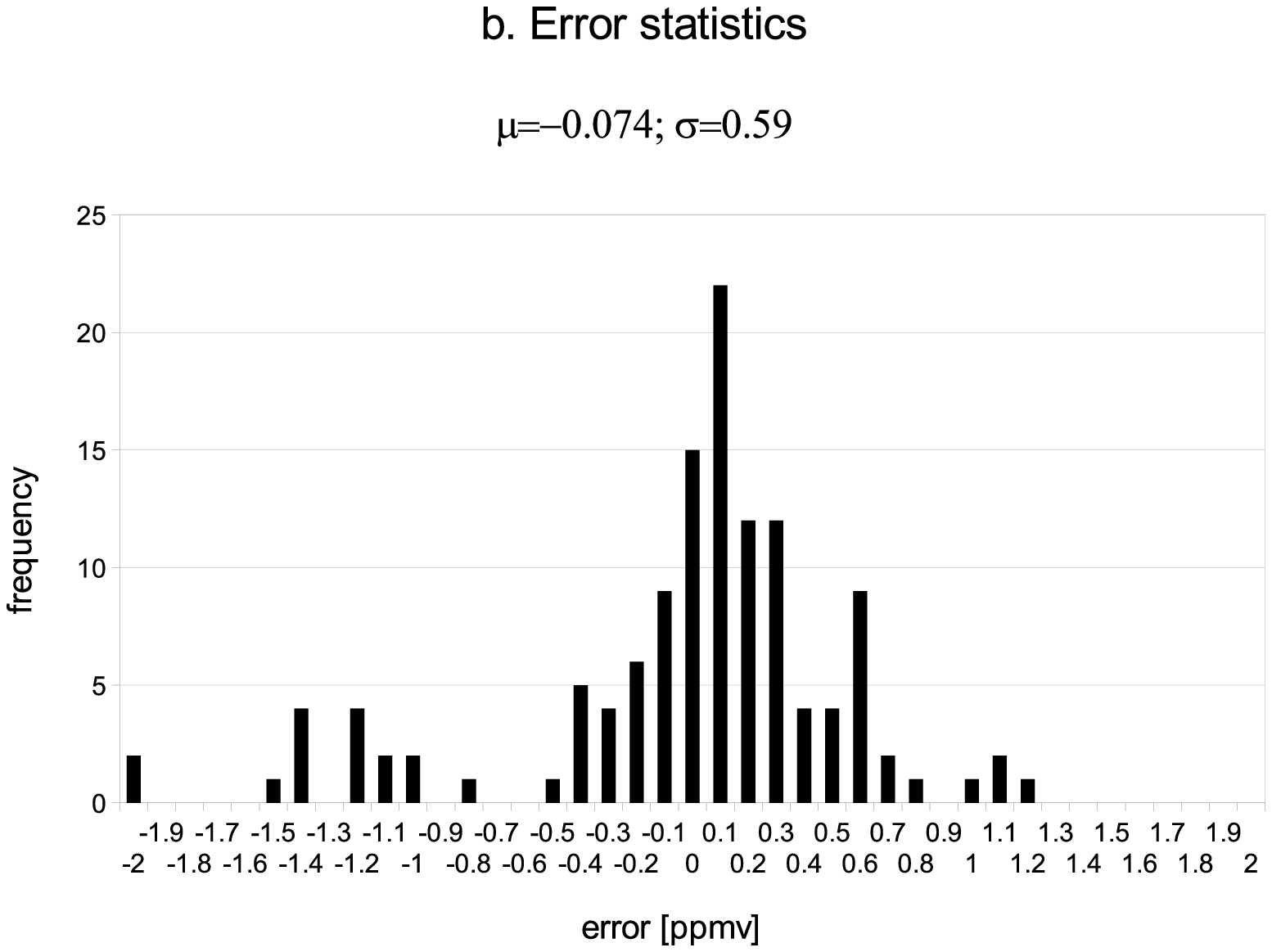}
  \caption{Validation of classic second-order proxy tracer ozone reconstruction against ozone sonde measurements. The econstruction was done globally on the 500 K isentrop from POAM III data between September 26, 1998 and November 17, 1998 using a measurement window of 2 days.}
  \label{global_classic_sonde}
\end{figure}

\begin{figure}
  \centering
  \includegraphics[width=0.9\textwidth]{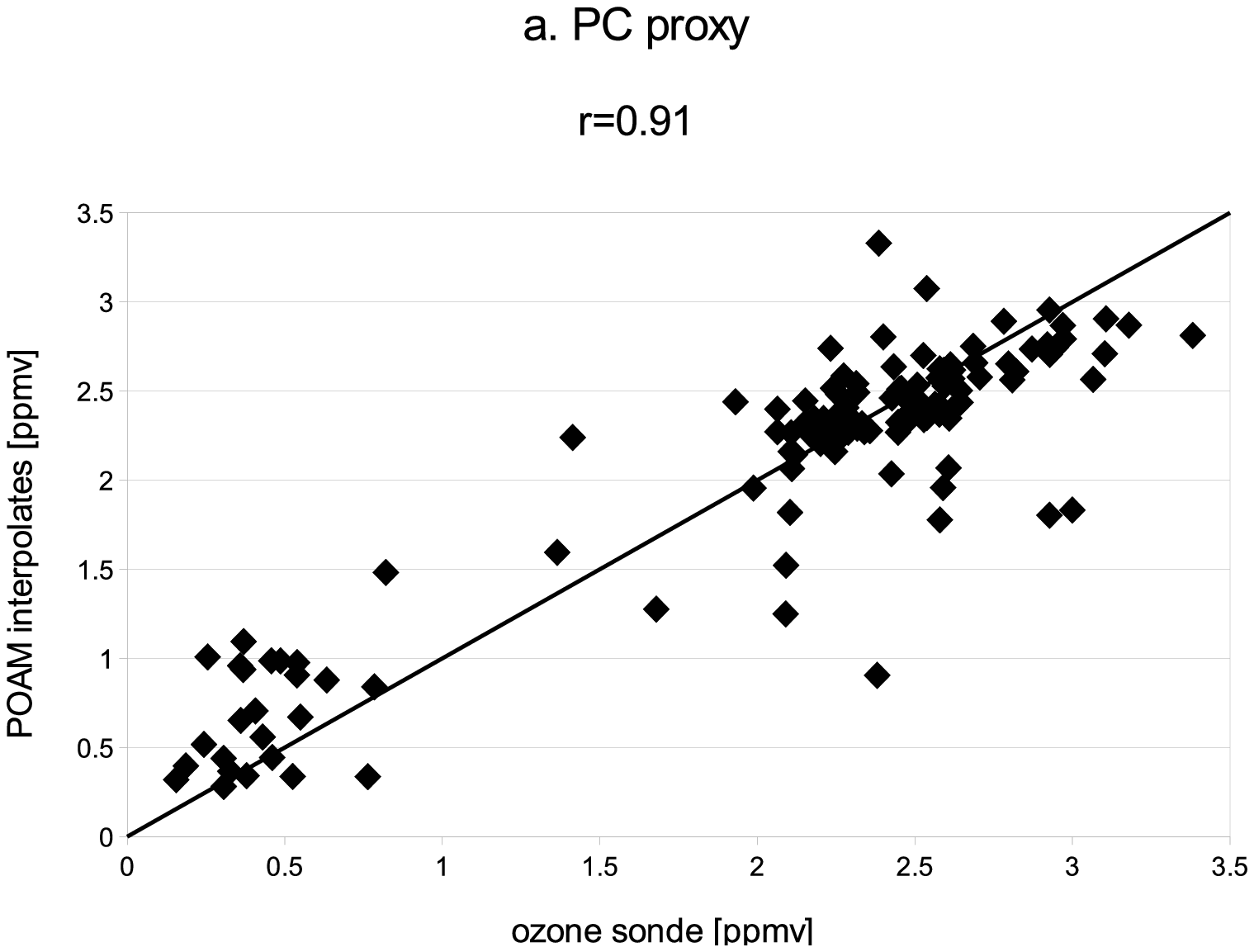}
  \includegraphics[width=0.9\textwidth]{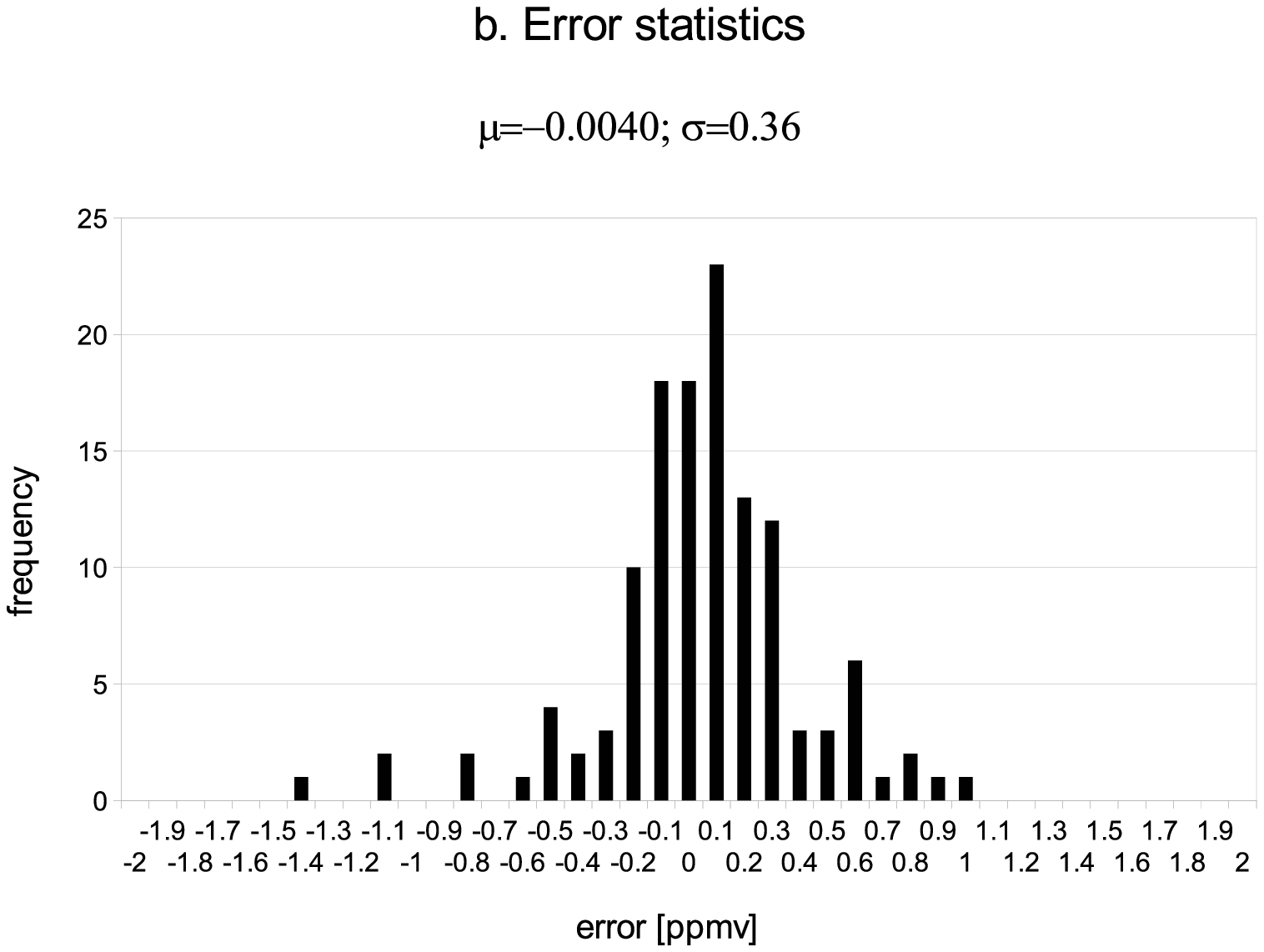}
  \caption{Validation of principal component proxy ozone reconstruction against ozone sonde measurements. The reconstruction was done globally on the 500 K isentrop from POAM III data between September 26, 1998 and November 17, 1998 using 5 principal components,  an integration time of 60 days, and a measurement window of 2 days.}
  \label{global_PC_sonde}
\end{figure}

In the first validation exercise, the POAM data was randomly separated into two
equal groups, each of which was used to predict the other.
Since the POAM measurements are closely grouped, falling into one of two narrow
latitude bands in the Arctic and Antarctic, and spaced at roughly 85 degrees
longitude between consecutive measurements, skill scores will tend to be
quite high.
PC proxy was compared to a classic proxy tracer with a second order fit.
The same tracer simulation as used for PC proxy was used to generate 
equivalent latitude fields \citep{Allen_Nakamura2003} for the classic proxy
tracer but with a two year spin-up.
Unlike in \citet{Randall_etal2002}, the reconstruction was done over the entire
globe.
Reconstructed ozone fields were linearly interpolated to match the locations
of the sonde measured test group.

Figure \ref{classic_cross_validation} shows the cross-validation results
for classic proxy tracer while Figure \ref{PC_cross_validation} shows
those for principal component proxy.
The correlation coefficient for the classic method was 0.977, as shown in 
the scatter plot in Figure \ref{classic_cross_validation}a.
The bias was -0.0063 parts-per-million by volume (ppmv) while the root-mean-square (RMS) error was 0.22 ppmv.
In the histogram error plot in Figure \ref{classic_cross_validation}b.
the errors have been normalized by the original error estimates for the
POAM III retrievals.
This makes it easy to compare residuals with the estimated errors
for the original ozone estimates.
When this is done, the bias becomes -0.023 while the RMS error is 1.18.
In other words, if the original error statistics are to be believed, 
the accuracy of ozone interpolates is almost as good as the original estimates
from the POAM satellite.

The correlation coefficient for the PC proxy method was 0.987, as shown in 
the scatter plot in Figure \ref{classic_cross_validation}a.
The bias was 0.0039 ppmv while the RMS error was 0.16 ppmv.
The normalized values are 0.021 for the bias and 0.86 for RMS error.
In other words, residuals for the reconstructed ozone are better,
on average, than the error bounds for the original retrievals!
On the other hand, the PC proxy method, while more accurate, appears to be
less stable as the pair of negative values in Figure \ref{PC_cross_validation}a
suggest.

\subsection{Sonde validation}

\begin{figure}
  \centering
  \begin{tabular}{cc}
  \includegraphics[width=0.45\textwidth]{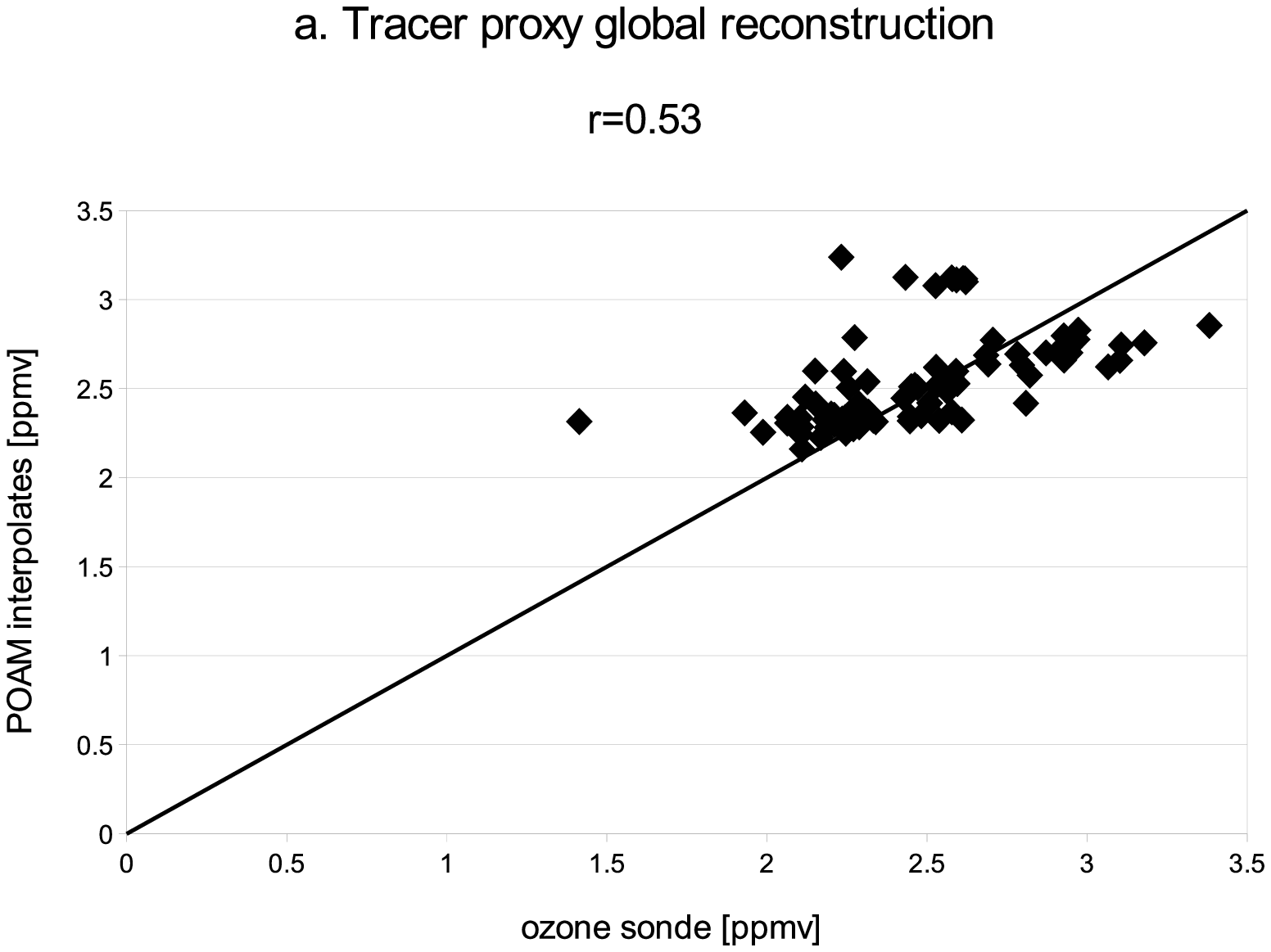} &
  \includegraphics[width=0.45\textwidth]{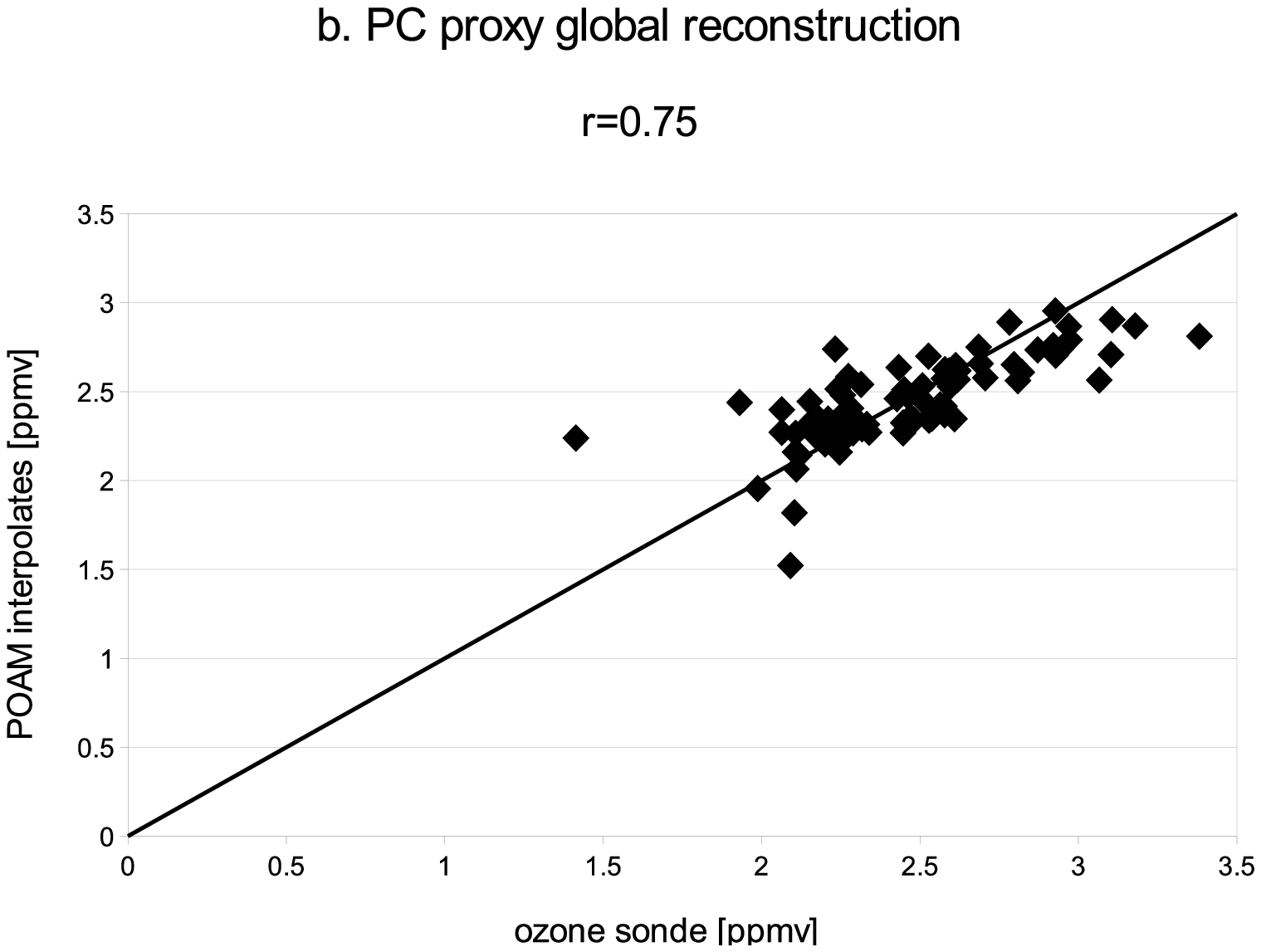} \\
  \includegraphics[width=0.45\textwidth]{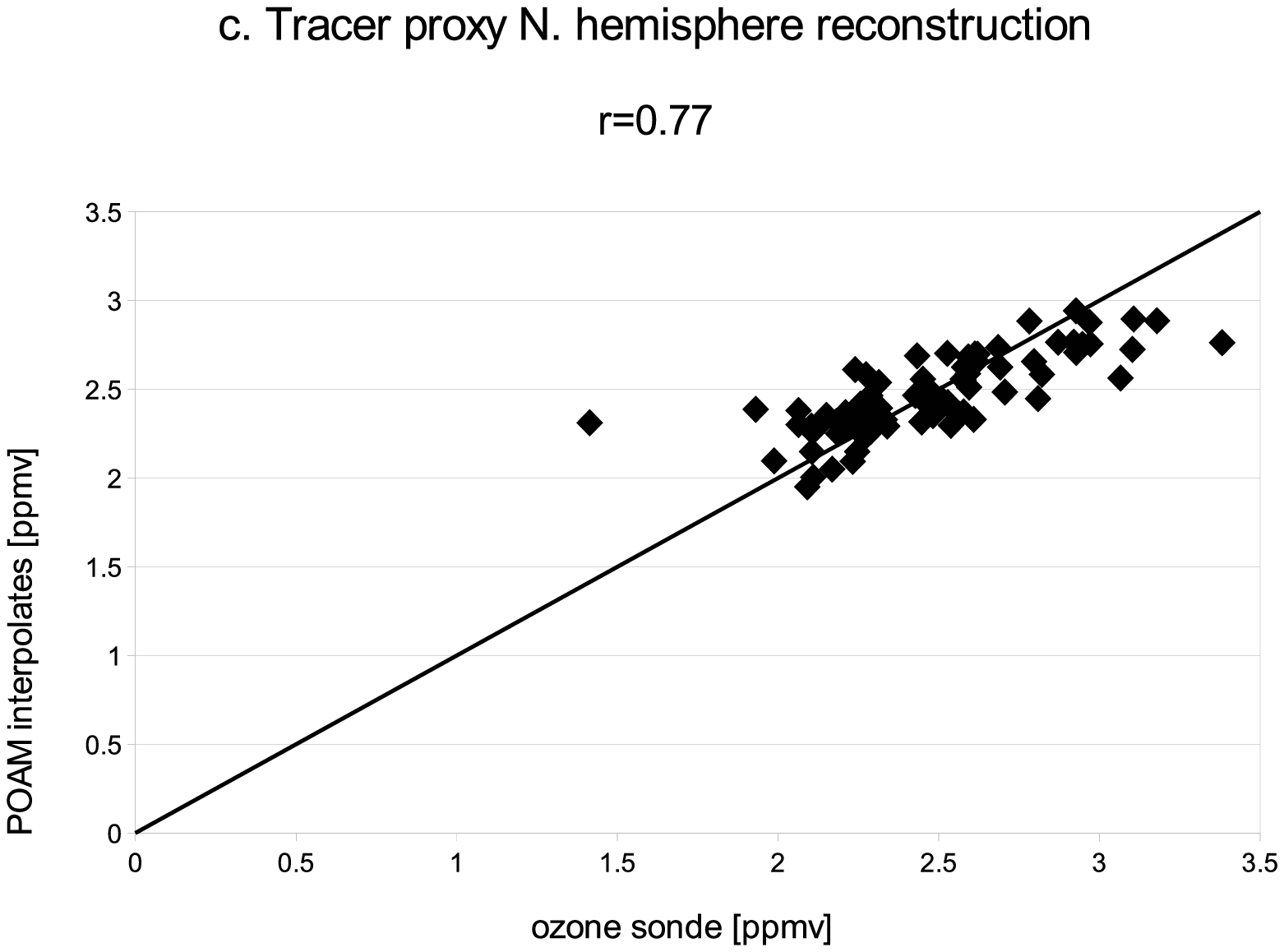} &
  \includegraphics[width=0.45\textwidth]{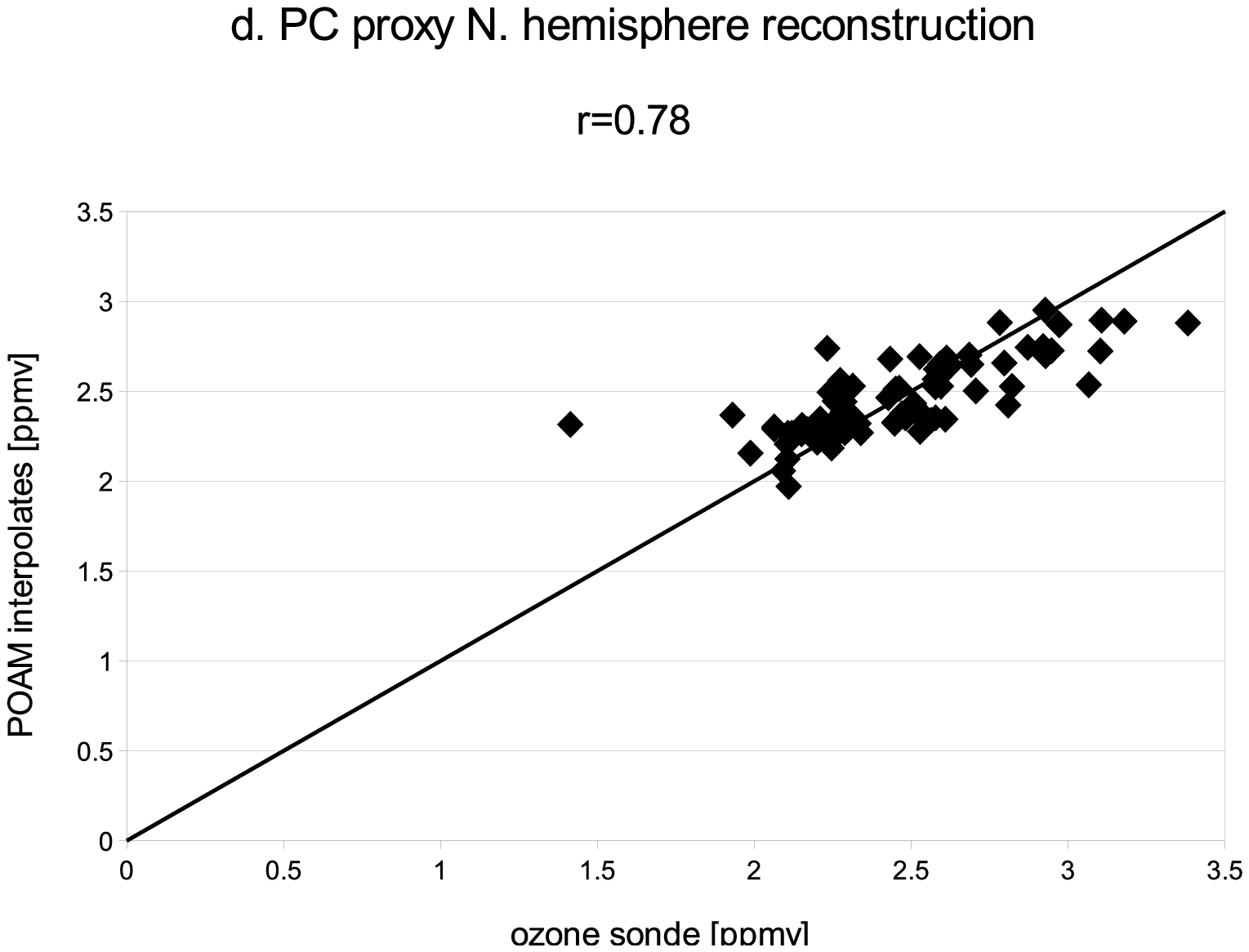}
  \end{tabular}
  \caption{Validation of ozone reconstruction against ozone sonde measurements
  in the Northern hemisphere: a. global reconstruction classic proxy tracer; b. global reconstruction PC proxy; c. classic proxy tracer Arctic POAM III only; d. PC proxy Arctic POAM III only.}
  \label{Nhemi}
\end{figure}

\begin{figure}
  \centering
  \begin{tabular}{cc}
  \includegraphics[width=0.45\textwidth]{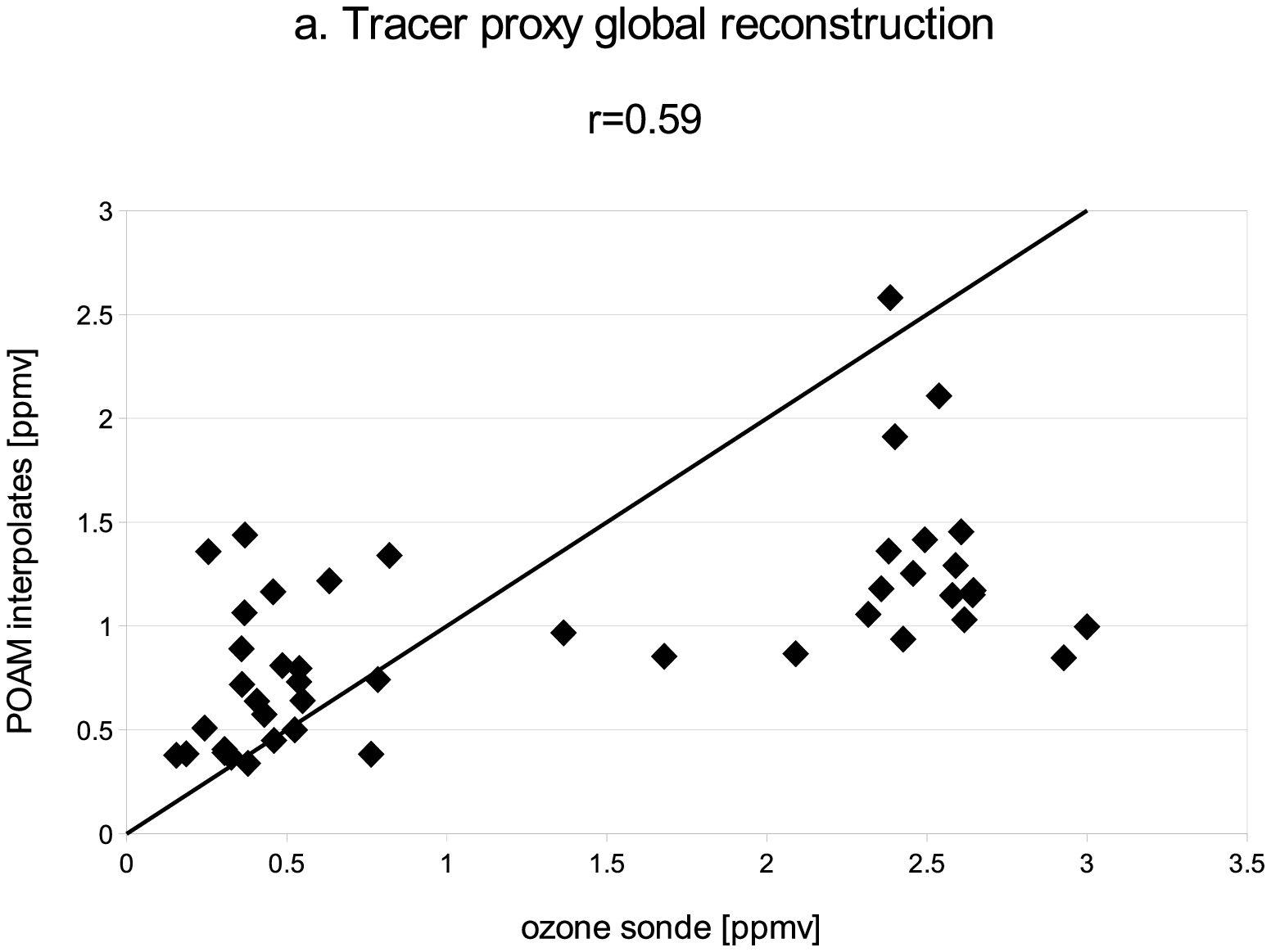} &
  \includegraphics[width=0.45\textwidth]{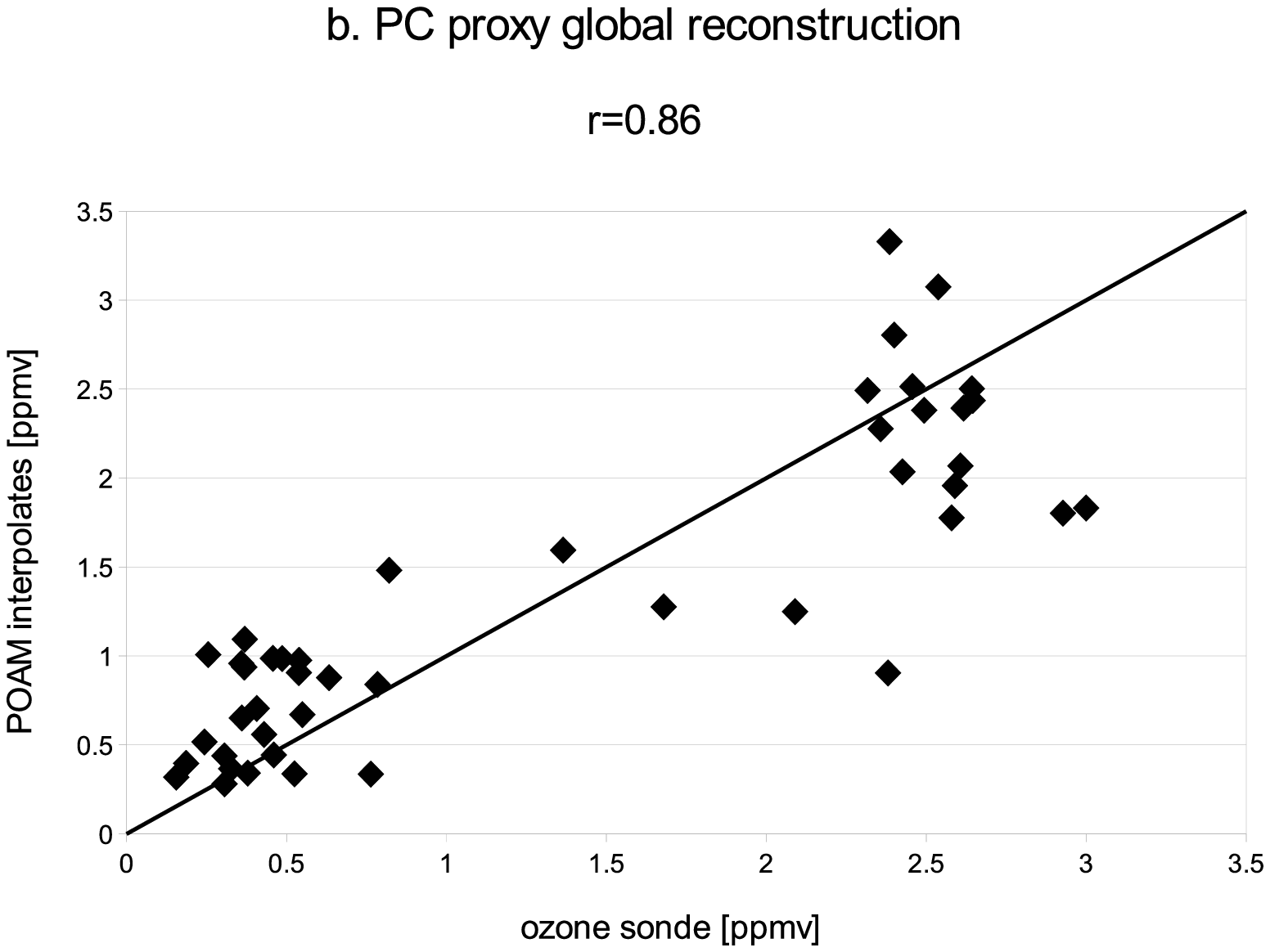} \\
  \includegraphics[width=0.45\textwidth]{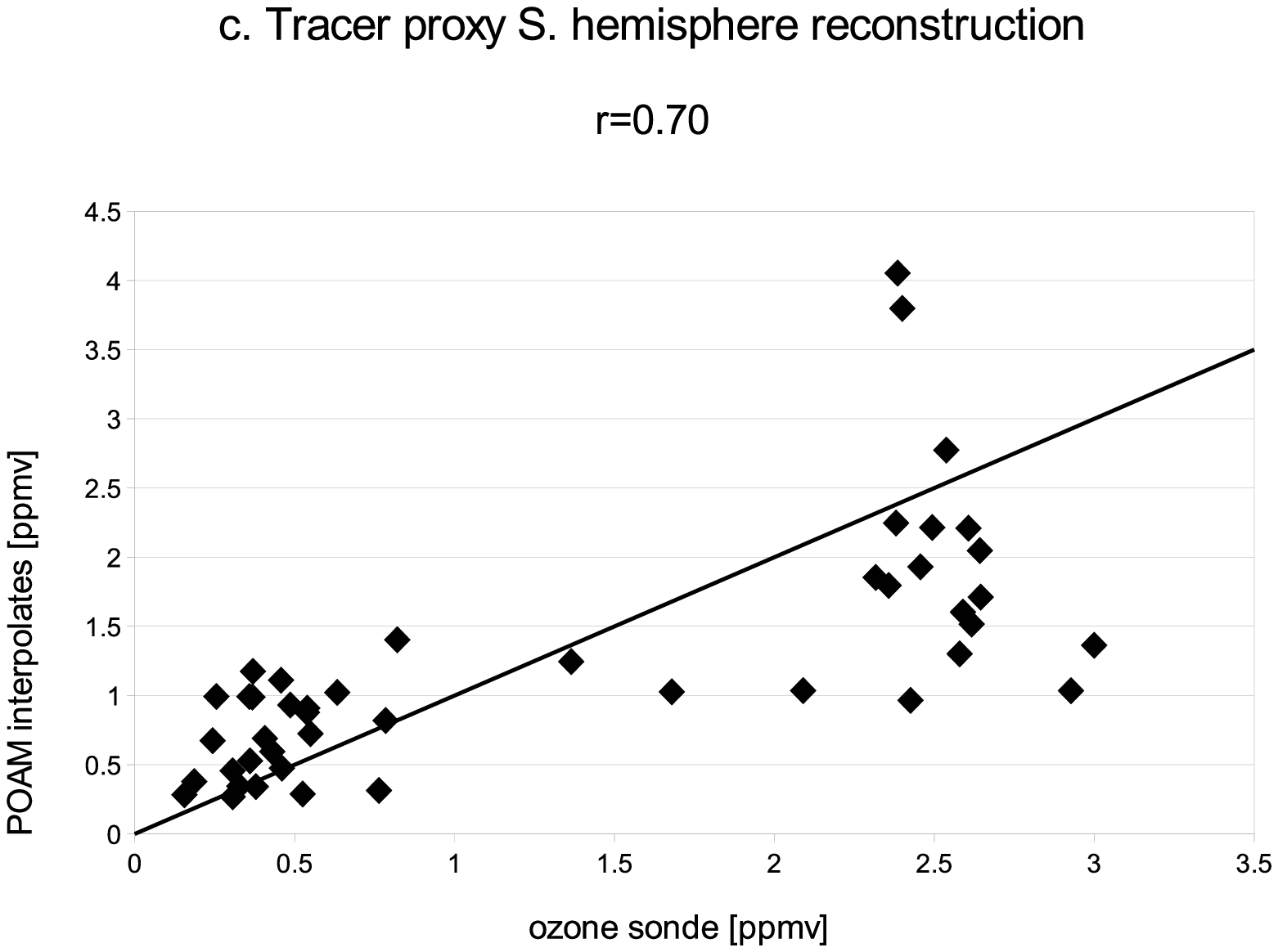} &
  \includegraphics[width=0.45\textwidth]{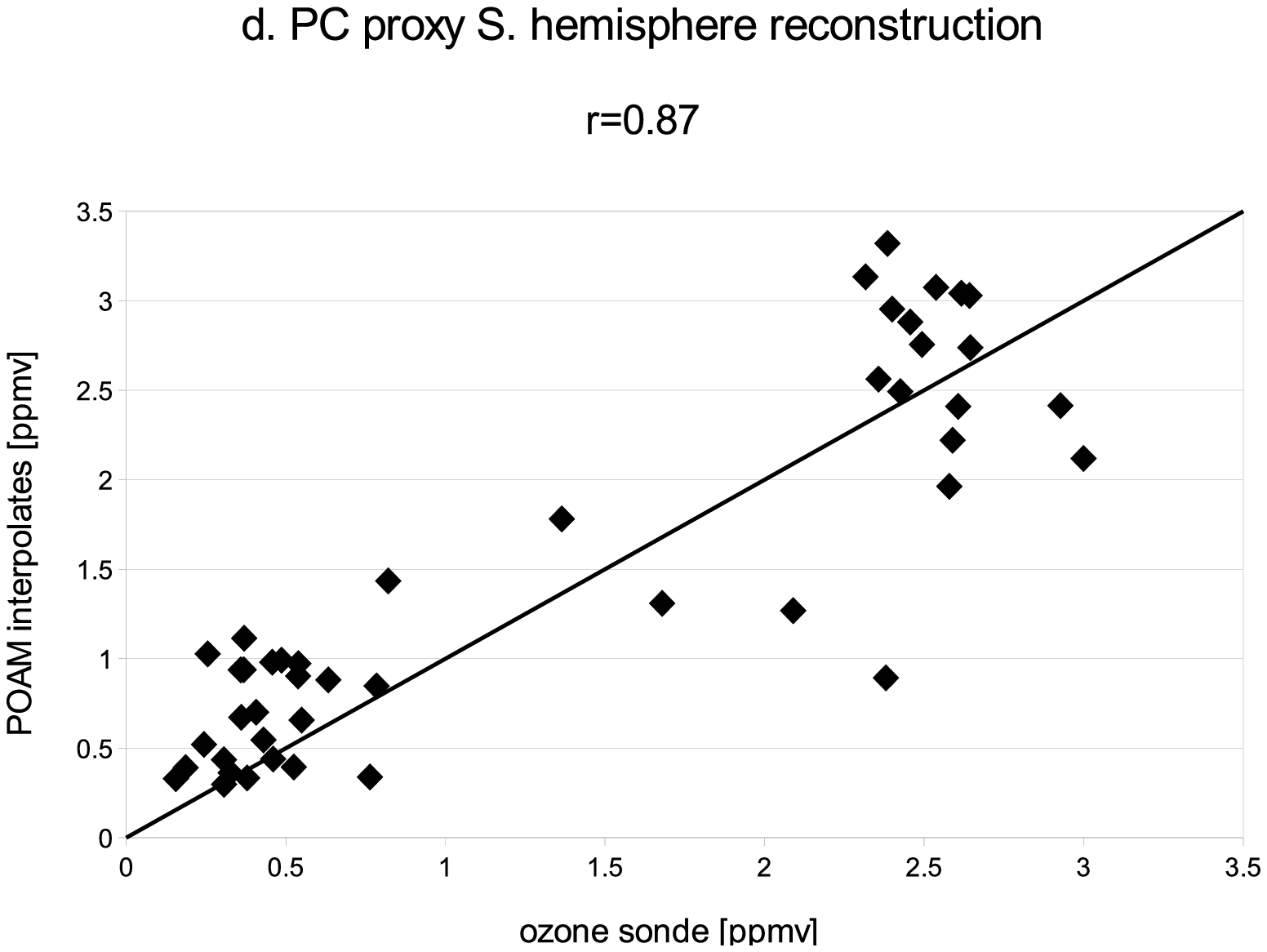}
  \end{tabular}
  \caption{Validation of ozone reconstruction against ozone sonde measurements
  in the Southern hemisphere: a. global reconstruction classic proxy tracer; b. global reconstruction PC proxy; c. classic proxy tracer Antarctic POAM III only; d. PC proxy Antarctic POAM III only.}
  \label{Shemi}
\end{figure}

\begin{table}
	\caption{Summary of ozone sonde trials. 
	$h_1$ is the hemisphere over which the reconstruction was done. 
$h_2$ is the hemisphere for which skill scores are shown.
-1 indicates Southern, +1 Northern and 0 the whole globe.
FAC2 is fraction of estimates within a factor of 2.}
	\label{summary}
	\begin{tabular}{|llll|lll|lll|}
	\hline
method & $k \lor N$ & $h_1$ & $h_2$ & $r$ & $\mu$ & $\sigma$ & $\mu/\sigma_q$ & $\sigma/\sigma_q$ & FAC2 \\\hline\hline
	Classic & 2 &  0 & 0 & 0.759 & -0.0742 & 0.592 & -0.0857 & 0.683 & 0.857 \\
	PC & 5 & 0 & 0 & 0.909 & -0.00396 & 0.362 & -0.00456 & 0.417 & 0.913 \\
	\hline
	Classic & 2 & 0 & +1 & 0.529 & 0.0782 & 0.296 & 0.235 & 0.891 & 1. \\
	PC & 5 & 0 & +1 & 0.748 & -0.0138 & 0.221 & -0.0415 & 0.664 & 1. \\
	\hline
	Classic & 2 & 0 & -1 & 0.590 & -0.349 & 0.846 & -0.337 & 0.817 & 0.6 \\
	PC & 5 &  0 & -1 & 0.858 & 0.0138 & 0.532 & 0.0133 & 0.514 & 0.756 \\
	\hline
	Classic & 2 & +1 & +1 & 0.772 & -0.00292 & 0.215 & -0.00877 & 0.645 & 1. \\
	PC & 2 & +1 & +1 & 0.780 & -0.00358 & 0.212 & -0.0107 & 0.638 & 1. \\
	\hline
	Classic & 1 & -1 & -1 & 0.703 & -0.0932 & 0.746 & -0.0901 & 0.721 & 0.733 \\
	PC & 2 & -1 & -1 & 0.887 & 0.140 & 0.485 & 0.136 & 0.469 & 0.756 \\
	\hline
\end{tabular}

\end{table}

In the second validation exercise, ozone fields were reconstructed from 
all available POAM data and compared with radio-sonde measurements.
Figure \ref{global_classic_sonde} shows the results for the classic proxy
tracer reconstruction validated against ozone sonde data from the WOUDC,
while figure \ref{global_PC_sonde} shows the same for the PC proxy method.
Correlation for the classic method stands at 0.76, with a bias of -0.074 ppmv
and a RMS error of 0.59 ppmv while the PC proxy method provides a correlation
coefficient of 0.91, a bias of -0.0040 ppmv and a RMS error of 0.36 ppmv.

For the two case studies discussed up to this point, 
the PC proxy method definitely has the edge. 
On the other hand, 
because ozone values tend to be higher in the Northern hemisphere than
they are in the South, skill scores may be unnaturally high.
As pointed out in the previous section, the original proxy tracer was 
typically only applied to one hemisphere at a time.
For a more well-rounded comparison, we should also apply them to only the
Northern or Southern hemisphere.
Meanwhile, to get a better idea of the skill of the methods, we should likewise
restrict comparisons to only a single hemisphere.

To get a more rigorous evaluation of each method we shall first confine them
to the Northern hemisphere
since ozone values there tend to fall in a narrower range of 
between 2 and 3 ppmv or so.
To this end, Figure \ref{Nhemi} shows scatter plots of only the
Northern hemisphere sonde launches as compared to the POAM interpolates.
In Figures \ref{Nhemi}a. and \ref{Nhemi}b. the reconstruction was done globally
but comparisons restricted to the Northern hemisphere.
Skill scores for the restricted comparisons show a correlation coefficient of 0.53,
a bias of 0.078 ppmv and a RMS error of 0.30 ppmv for the classic proxy
and a correlation coefficient of 0.75, a bias of -0.014 and a RMS error of 0.22 for
the PC proxy.
Figures \ref{Nhemi}c. and \ref{Nhemi}d. show results for interpolates for which the
reconstruction was done in the Northern hemisphere using 
Arctic POAM III measurements only.
Scores improve considerably for the classic proxy tracer and now almost
match PC proxy.
Skills scores for the Northern Hemisphere reconstruction show a correlation coefficient
of 0.77, a bias of -0.0029 ppmv, and a RMS error 0.22 ppmv for the classic
tracer proxy and a correlation coefficient of 0.78, a bias of -0.0035 ppmv,
and a RMS error of 0.21 ppmv for the PC proxy.
For the N. hemisphere reconstruction, only two (2) principal components were
used in PC proxy.

Results for the Southern hemisphere, shown in Figure \ref{Shemi}, are far more 
favourable for the PC proxy method.
For global reconstruction restricted to the S. hemisphere, classic proxy
returned a correlation coefficient of 0.59, a bias of -0.35 ppmv and a RMS error 
of 0.85 ppmv while PC proxy gave a correlation coefficient of 0.86, a bias of 0.014
and a RMS error of 0.53.
For S. hemisphere reconstruction, the numbers are: a correlation coefficient of 0.70,
a bias of -0.0093 ppmv and a RMS error of 0.75 ppmv for classic proxy
while PC returned a correlation coefficient of 0.89, a bias of 0.14 and a RMS error
of 0.49.
Once again, only two principal components were used in the hemispherical PC
proxy reconstruction.
For the classic proxy tracer, hemispherical interpolates were to only first-, rather than
second-order ($N=1$).

All results for the sonde validation are summarized in Table \ref{summary},
including relative bias, normalized RMS error and FAC2. 
Both relative bias and normalized RMS are normalized by the standard 
deviation of the measurement data.
FAC2 is the fraction of estimates that are within a factor of 2 of the 
measurement values.

\section{Discussion and conclusions}

The raw skill scores listed in the previous section show the PC proxy method
to be more accurate than the classic proxy tracer method. 
Differences are largest when the reconstruction is performed globally.
When the reconstruction is performed over only a single hemisphere,
differences are smaller but nonetheless significant.
The two methods are almost the same in the Northern hemisphere but in 
the Southern hemisphere PC proxy is superior even when the PC proxy
reconstruction is performed globally while the classic proxy tracer is 
performed only in the Southern hemisphere.

We surmise that part of the advantage lies in having more degrees of freedom
or at least (in the case of $k \le N+1$), more meaningful degrees of freedom:
more than one possible tracer configuration is represented.
Moreover, because it is using information from shorter time scales, it
is better able to fit more active tracers that are changing more rapidly,
as ozone would be in the Southern hemisphere during winter time.

Unfortunately, because of these extra degrees of freedom, the method can also
be more unstable.
This is observed in negative values that showed up first in the cross-validation
exercise in Section \ref{cross_validation}, but also in ringing and regions
of negative concentration towards the equator in some of the reconstructed
fields.
This is to be expected: the POAM III data covers only a very small latitude 
band so that values derived near the equator could be considered an
extrapolation and are vulnerable to ``over-fitting''.
Such instabilities also appeared in the classic method, although less often.
They don't affect the validation results so much because most of the 
radiosonde launch stations are in the higher latitudes, as seen in
Figure \ref{ozone_sonde_stations}.

In PC proxy, instabilities can be reduced by using fewer principal components
which reduces the number of degrees of freedom.
Unfortunately, this also tends to reduce the accuracy in the areas where the
method does not fail.
A good example of this phenomenon is the classic proxy in the Southern
hemisphere. Here we have used only a first-order fit because while a 
second-order fit produced higher correlation scores, it also returned a
much larger bias and RMS error.
Stability is traded off for greater potential precision.
The best remedy would be either to have more measurements sampling more of
the globe or to simply not use interpolates far outside the range of the
measurements.

The ability to accurately combine information over the whole globe
would appear to be the biggest advantage of the method.
As compared to the earlier method, PC proxy
allows for seamless reconstruction even across the equator.
Note that there is very little difference between the global reconstructions
and those restricted to a single hemisphere, especially in the Southern
hemisphere.
Even though there isn't a lot of cross-talk between the two hemispheres, the
the greater degrees of freedom in the PC proxy method overcomes this.
It would be instructive to test the method on an instrument that samples
more broadly and using more principal components.

Principal component proxy tracer reconstruction is shown to be a powerful
technique that 
has many of the advantages of prognostic assimilation models but with
less of the associated complexity.
In particular, there is no need for explicitly modelled sources and sinks.
It has the advantage over the classic proxy tracer method because it takes
into account more than one possible configuration of a passive tracer, operates over
shorter time scales and has more parameters to tune for optimal accuracy
and stability.
It is also more accurate when performed over the entire Earth rather than
a single hemisphere, allowing for seamless reconstruction of global fields.
It would be straightforward to adapt to three dimensional reconstruction 
as well as reconstruction that relies on column measurements or even un-inverted
level 1 satellite measurements rather than point measurements.
Consider the vector, $\vect w_i$, in Equation (\ref{PC_proxy_def})
to be a set of weights for integrating a column of air or performing
a radiative transfer simulation rather than a set of interpolation
coefficients.

\section{Acknowledgements}

Thanks to the National Center for Environmental Prediction and the National Center for Atmospheric Research for the reanalysis data used in the simulations.
Thanks also to World Ozone and Ultraviolet Data Center and Environment Canada for ozone sonde data.
And thanks especially to my former colleagues at the Naval Research Laboratory for POAM III ozone data.

Contour maps were created with Generic Mapping Tools (GMT) while scatter plots
and historgrams were done in Open Office.

\addcontentsline{toc}{section}{References}

\bibliography{proxy2.bib}

\appendix

\section{Model properties}

\subsection{Mass conservation}

\label{mass_conservation_derivation}

Suppose that:
\begin{equation}
	\sum_i q_i = const.
	\label{constant_mass}
\end{equation}
This will be true for non-divergent flows on equal area grids.
Then:
\begin{eqnarray}
	\sum_i \sum_j r_{ij} q_j & = & \sum_j q_j \\
	\sum_j q_j \left ( \sum_i r_{ij} - 1 \right ) & = & 0
\end{eqnarray}
Therefore:
\begin{equation}
	\sum_i r_{ij} = 1
\end{equation}
If (\ref{constant_mass}) is true, then:
\begin{equation}
	\frac{\mathrm d}{\mathrm d t}\sum_i q_i = 0
\end{equation}
is also be true. Continuing:
\begin{eqnarray}
	\sum_i \frac{\mathrm d q_i}{\mathrm d t} & = & 0 \\
\sum_i \sum_j a_{ij} q_j & = & 0 \\
\sum_j q_j \sum_i a_{ij} & = & 0
\end{eqnarray}
which shows the second part of (\ref{columns_sum_to_one}) and 
(\ref{columns_sum_to_zero}):
\begin{equation}
	\sum_i a_{ij} = 0
\end{equation}

\subsection{Diffusion and the Lyapunov spectrum}

\label{Lyapunov_exponents_less_than_zero}

A discrete tracer mapping will always require some 
amount of diffusion.  This means that the tracer configuration will 
tend towards a uniform distribution over time, 
that is, it will ``flatten out.''  We can
show that, given the constraint in (\ref{constant_mass}), 
a tracer field with all the same values has the smallest magnitude.  
Suppose there are only two elements in the 
tracer vector, $\vect q=\lbrace q,~q \rbrace$.  The magnitude of the vector is:
\begin{equation}
|\vect q|=\sqrt{q^2+q^2}=\sqrt{2} q
\end{equation}
Now we introduce a separation between the elements, $2\Delta q$, that 
nonetheless keeps the sum of the elements constant:
\begin{eqnarray}
|q+\Delta q,~q-\Delta q| & = & \sqrt{(q+\Delta q)^2+(q-\Delta q)^2} \\
& = & \sqrt{2}\sqrt{q^2+(\Delta q)^2} \ge \sqrt{2} q
\end{eqnarray}
This will generalize to higher-dimensional vectors.  In general, we can
say that:
\begin{equation}
\vect q R^T R \vect q \le | \vect q |^2
\label{tracer_map_inequality}
\end{equation}
Implying that for the eigenvalue problem,
\begin{eqnarray}
R^T R \vect v & = & s^2 \vect v \nonumber\\
s^2 & \le & 1 \label{SV_inequality}
\end{eqnarray}
Therefore the Lyapunov exponents are all
either zero or negative.
Note however that this does not constitute a proof; the actual proof is more 
involved.

To prove (\ref{SV_inequality}) from (\ref{tracer_map_inequality}), we first
expand $\vect q$ in terms of the right singular values, 
$\lbrace \vect v_i \rbrace$:
\begin{equation}
	\vect q = \sum_i c_i \vect v_i
\end{equation}
where $\lbrace c_i \rbrace$ are a set of coefficients.
Substituting this into the left-hand-side of (\ref{tracer_map_inequality}):
\begin{eqnarray}
	\vect q R^T R \vect q & = & \sum_i c_i \vect v_i \sum_i c_i s_i^2 \vect v_i \\
   & = & \sum_i \sum_j c_i c_j s_i^2 \vect v_i \vect v_j \\
   & = & \sum_i \sum_j c_i c_j s_i^2 \delta_{ij} \\
	  & = & \sum_i c_i^2 s_i^2
\end{eqnarray}
where $\delta$ is the Kronecker delta.
Similarly, we can show that:
\begin{equation}
	\vect q \cdot \vect q = \sum_i c_i^2
\end{equation}
If we assume that $s_i \le 1$ for every $i$, then:
\begin{equation}
	\sum_i c_i^2 s_i^2 \le \sum_i c_i^2 
	\label{diffusive_inequality_in_terms_of_SVs}
\end{equation}
since each term on the left side is less-than-or-equal-to the
corresponding term on the right side. 
Note that in order for the inequality in 
(\ref{diffusive_inequality_in_terms_of_SVs}) to be broken, at least one
singular value must be greater-than one.
Therefore (\ref{tracer_map_inequality}) is true for every $\vect q$
if-and-only-if (\ref{SV_inequality}) is true for every $s$.
In the language of set theory and first-order logic:
\begin{equation}
	\forall \vect q \in \Re^n ~ (| R \vect q |^2 \le |\vect q |^2) \iff \forall s \in \Re | ~R^T R \vect v = s^2 \vect v ~ (s \le 1)
	\label{theorem}
\end{equation}

\section{Deviation from equal area}

\label{equal_area}

Here we calculate the ratio between the largest and smallest grid boxes in the
azimuthal equidistant coordinate system.
First we show that there is no distortion at the pole:
\begin{eqnarray}
	\lim_{x\rightarrow 0, ~y\rightarrow 0} \left (\frac{\mathrm ds}{\mathrm d x} \right )^2 & = & 
	\lim_{x\rightarrow 0, ~y\rightarrow 0} \left (\frac{\mathrm ds}{\mathrm d y} \right )^2  \\
	& = & \lim_{x\rightarrow 0, ~y\rightarrow 0} \frac{1}{r^2} \left [
	\frac{R_E^2}{r^2} \sin^2 \left (\frac{r}{R_E} \right ) y^2 + x^2 \right ] \\
 & = & \frac{1}{r^2} \left [\frac{R_E^2}{r^2} \left (\frac{r}{R_E} \right )^2
	y^2 + x^2 \right ] \\ 
	& = & 1
\end{eqnarray}
hence the ratio between projected and unprojected areas is 1.
Grid areas become progressively smaller the further from the pole you get.
Since the projection is hemi-spherical, $r$ takes on a maximum value at the
equator:
\begin{equation}
	r=\pi R_E/2
\end{equation}
Hence the largest possible values for $x$ and $y$ are:
\begin{equation}
	x = y = \frac{\pi R_E}{2 \sqrt{2}}
\end{equation}
which represents a point on the equator along a diagonal from the origin 
in the projected coordinate system.
The metric coefficients can be calculated:
\begin{eqnarray}
	\left (\frac{\mathrm ds}{\mathrm d x} \right )^2 & = & 
	\left (\frac{\mathrm ds}{\mathrm d y} \right )^2 \\
     & = & \frac{4}{\pi^2 R_e^2} \left [
	\frac{4 R_E^2}{\pi^2 R_E^2} \sin^2 \left (\frac{\pi R_E}{2 R_E} \right ) 
\frac{\pi^2 R_E^2}{8} + \frac{\pi^2 R_E^2}{8} \right ] \\
& = & \frac{1}{\pi^2 R_E^2} \left ( 2 R_E^2 + \frac{\pi^2 R_E^2}{2} \right ) \\
& = & \frac{2}{\pi^2} + \frac{1}{2} \\
& \approx & 0.703
\end{eqnarray}

\end{document}